\documentstyle[12pt,epsf,epsfig]{article}
\newcount\timecount
\newcount\hours \newcount\minutes  \newcount\temp \newcount\pmhours
\hours = \time
\divide\hours by 60
\temp = \hours
\multiply\temp by 60
\minutes = \time
\advance\minutes by -\temp
\def\hour{\the\hours}
\def\minute{\ifnum\minutes<10 0\the\minutes
            \else\the\minutes\fi}
\def\clock{
\ifnum\hours=0 12:\minute\ AM
\else\ifnum\hours<12 \hour:\minute\ AM
      \else\ifnum\hours=12 12:\minute\ PM
            \else\ifnum\hours>12
                 \pmhours=\hours
                 \advance\pmhours by -12
                 \the\pmhours:\minute\ PM
                 \fi
            \fi
      \fi
\fi
}

\def\monthname{\relax\ifcase\month 0/\or January\or February\or
   March\or April\or May\or June\or July\or August\or September\or
   October\or November\or December\else\number\month/\fi}

\def\bold#1{\setbox0=\hbox{$#1$}%
     \kern-.025em\copy0\kern-\wd0
     \kern.05em\copy0\kern-\wd0
     \kern-.025em\raise.0433em\box0 }

\def\beq{\begin{equation}}
\def\eeq{\end{equation}}
\def\ba{\begin{eqnarray}}
\def\ea{\end{eqnarray}}

\def\ss{\scriptscriptstyle}
\def\ga{\mathrel{\raise.3ex\hbox{$>$\kern-.75em\lower1ex\hbox{$\sim$}}}}
\def\la{\mathrel{\raise.3ex\hbox{$<$\kern-.75em\lower1ex\hbox{$\sim$}}}}
\def\gev{{\rm \, Ge\kern-0.125em V}}
\def\tev{{\rm \, Te\kern-0.125em V}}
\def\gyr{{\rm \, G\kern-0.125em yr}}

\def\gappeq{\mathrel{\rlap {\raise.5ex\hbox{$>$}}
{\lower.5ex\hbox{$\sim$}}}}
\def\lappeq{\mathrel{\rlap{\raise.5ex\hbox{$<$}}
{\lower.5ex\hbox{$\sim$}}}}
\def\Toprel#1\over#2{\mathrel{\mathop{#2}\limits^{#1}}}

\def\m12{m_{1\!/2}}

\def\mz{m_{\ss Z}}

\begin{document}
\begin{titlepage}
\pagestyle{empty}
\baselineskip=21pt
\rightline{\tt hep-ph/0405110}
\rightline{CERN--TH/2004-069, UMN--TH--2307/04, FTPI--MINN--04/18}
\vskip 0.2in
\begin{center}
{\large {\bf Very Constrained Minimal Supersymmetric Standard Models}}
\end{center}
\begin{center}
\vskip 0.2in
{\bf John~Ellis}$^1$, {\bf Keith~A.~Olive}$^{2}$, {\bf Yudi Santoso}$^{2}$
and {\bf Vassilis~C. Spanos}$^{2}$
\vskip 0.1in
{\it
$^1${TH Division, CERN, Geneva, Switzerland}\\
$^2${William I. Fine Theoretical Physics Institute, \\
University of Minnesota, Minneapolis, MN 55455, USA}}\\
\vskip 0.2in
{\bf Abstract}
\end{center}
\baselineskip=18pt \noindent

We consider very constrained versions of the minimal supersymmetric
extension of the Standard Model (VCMSSMs) which, in addition to
constraining the scalar masses $m_0$ and gaugino masses $m_{1/2}$ to be
universal at some input scale, impose relations between the trilinear and
bilinear soft supersymmetry breaking parameters $A_0$ and $B_0$. These
relations may be linear, as in simple minimal supergravity models, or
nonlinear, as in the Giudice-Masiero mechanism for generating the
Higgs-mixing $\mu$ term. We discuss the application of the electroweak
vacuum conditions in VCMSSMs, which may be used to make a prediction for
$\tan \beta$ as a function of $m_0$ and $m_{1/2}$ that is usually unique.
We baseline the discussion of the parameter spaces allowed in VCMSSMs by
updating the parameter space allowed in the CMSSM for fixed values of
$\tan \beta$ with no relation between $A_0$ and $B_0$ assumed {\it a
priori}, displaying contours of $B_0$ for a fixed input value of $A_0$,
incorporating the latest CDF/D0 measurement of $m_t$ and the latest BNL
measurement of $g_\mu - 2$. We emphasize that phenomenological studies of
the CMSSM are frequently not applicable to specific VCMSSMs, notably those
based on minimal supergravity, which require $m_0 = m_{3/2}$ as well as
$A_0 = B_0 + m_0$. We then display $(m_{1/2}, m_0)$ planes for selected
VCMSSMs, treating in a unified way the parameter regions where either a
neutralino or the gravitino is the LSP. In particular, we examine in
detail the allowed parameter space for the Giudice-Masiero model.

\vfill
\leftline{CERN--TH/2004-069}
\leftline{May 2004}
\end{titlepage}
\baselineskip=18pt

\section{Introduction}

Supersymmetry is one of the most appealing extensions of the Standard
Model (SM), for many reasons including the hierarchy problem, its
necessity in string theory, unification of the SM gauge couplings, the
suggestion of a light Higgs boson, the possibility that the astrophysical
cold dark matter might be provided by the lightest supersymmetric particle
(LSP) and (just possibly)  the anomalous magnetic moment of the muon,
$g_\mu - 2$. However, supersymmetry is a general framework that
accommodates many new degrees of freedom. The simplest possible
realization of supersymmetry is the minimal supersymmetric extension of
the SM (MSSM). Four types of supersymmetry-breaking parameters appear in
the MSSM: scalar masses $m_0$, gaugino masses $m_{1/2}$, trilinear
couplings $A$ and a bilinear coupling $B$ in the Higgs sector. In the MSSM
alone, the number of free parameters associated with soft supersymmetry
breaking exceeds 100, unless one assumes some degree of universality for
the sparticles with different quantum numbers and flavours. In
phenomenological studies of supersymmetry, the values of $m_0$ for the
different sflavours are often constrained to be universal at some input
GUT scale, as are the values of $m_{1/2}$ for the different SM gauge group
factors, and the $A$ parameters corresponding to different SM Yukawa
couplings, a framework often called the CMSSM.

One may go even further, and assume some relation(s) between the
parameters $m_0, m_{1/2}$, $A$, $B$ and the gravitino mass $m_{3/2}$. In 
particular, many very constrained
versions of the MSSM (VCMSSMs) derive or postulate relations between the
$A$ and $B$ parameters, which we parametrize as $A \equiv {\hat A} m_0, B
\equiv {\hat B} m_0$. These relations may be linear: for example, generic
minimal supergravity models predict that ${\hat B} = {\hat A} - 
1$~\cite{BIM,mark} as well as $m_0 = m_{3/2}$, and
the simplest Polonyi model \cite{pol} of supersymmetry breaking further 
predicts that
$\vert {\hat A} \vert = 3 - \sqrt{3}$~\cite{bfs}. On the other hand, a 
prominent example of a nonlinear relation is 
\beq
\hat{B} = \frac{2 \hat{A} - 3}{\hat{A} - 3},
\label{GMrelation}
\eeq
which appears in the Giudice-Masiero mechanism~\cite{gm} for generating 
the $\mu$ term.

In the CMSSM, one may regard $m_0, m_{1/2}$ and $A$ as independent
parameters, and use the two electroweak vacuum conditions resulting from
the specification of $m_Z$ and the ratio of Higgs vacuum expectation
values, $\tan \beta$, to fix $|\mu|$ and the pseudoscalar Higgs mass
$m_A$, which is equivalent to fixing ${\hat B}$. As we show in this paper
with some explicit examples, the value of ${\hat B}$ that results for any
given choice of $m_0, m_{1/2}$ and ${\hat A}$ may not correspond to any
plausible theoretical model. Conversely, in a VCMSSM where ${\hat B}$ is 
fixed in terms of ${\hat A}$, one can use the electroweak vacuum
conditions to predict $\tan \beta$ as a function of $m_0, m_{1/2}$ and
${\hat A}$. In a previous paper~\cite{AB1}, we demonstrated this type of 
prediction
for a few specific VCMSSMs with linear relations between ${\hat A}$ and
${\hat B}$, including minimal supergravity, with the simplest Polonyi
model as a special case.

In this paper, we extend the previous discussion to include the 
Giudice-Masiero model. In this case, in addition to the relation
(\ref{GMrelation}) between $\hat B$ and $\hat A$, 
the value of $\mu$ is in principle 
also predicted as
\beq
\left| \frac{\mu}{m_0} \right| = \left| \lambda \frac{\hat{A} - 
3}{\sqrt{3}}  
\right|
\label{GMlambda}
\eeq
where $\lambda/M_{Pl}$ is the coupling between a hidden sector superfield and
the two Higgs superfields. The value of $\lambda$ is presumably not completely
arbitrary: for example, one should probably require $|\lambda| \lappeq
O(1)$. This bound on $|\lambda|$ in turn imposes a range on the ratio $\mu
/ m_0$ for a given $\hat A$.  Since the value of $|\mu|$ is an output quantity
in our approach
to VCMSSMs, one must check that $|\lambda|$ is not very large,
which could in principle restrict the ranges of the input parameters.

The first step in this paper is to discuss the application of the
electroweak vacuum conditions. In principle, more than one value of $\tan
\beta$ might be consistent with a given VCMSSM for some specific choice of
$(m_0, m_{1/2})$ \cite{dn1}. In practice, over large regions of the $(m_0,
m_{1/2})$ we find only one solution for $\tan \beta$, as we explain in
some detail.  We also discuss the renormalization of the input relation
between $A$ and $B$ in a generic VCMSSM, including the relation between
the input and electroweak-scale values of $B$ and the one-loop threshold
corrections at the electroweak scale.

The second step is to update previous analyses of the CMSSM, including
some updates in calculations of the supersymmetric particle spectrum, as well
as the latest information on $m_t$, $g_\mu - 2$, $b \to s \gamma$ and $B_s
\to \mu^+ \mu^-$. We demonstrate that the values of ${\hat B}$ required
in the CMSSM for generic values of $\tan \beta$ and ${\hat A}$ do not fit
within favoured VCMSSM frameworks, such as those based on minimal 
supergravity or the Giudice-Masiero model.

We then discuss the $(m_0, m_{1/2})$ planes for some specific VCMSSMs, 
taking into account the fact that
minimal supergravity models predict that $m_0 = m_{3/2}$ before 
renormalization, which is not necessarily the case 
in a
generic CMSSM. This relation enables one to delineate the regions where
the LSP is the lightest neutralino $\chi$, the lighter ${\tilde \tau}$ or
the gravitino ${\tilde G}$. We present unified descriptions of the $\chi$
and ${\tilde G}$ LSP regions for some specific VCMSSMs, incorporating the
constraints on decays of the next-to-lightest supersymmetric particle
(NSP) into a gravitino LSP that are imposed by concordance between the
Big-Bang nucleosynthesis (BBN) and cosmological microwave background (CMB)
determinations of the baryon-to-entropy ratio~\cite{GDM,feng,CEFO,had}. 
Finally, we discuss the
Giudice-Masiero model in more detail, finding that the implied values of
$|\lambda|$ in allowed regions of parameter space are generally $\gappeq
O(1)$, particularly in the gravitino LSP region.

\section{Summary of Models of Supersymmetry Breaking}

As discussed in~\cite{AB1}, we assume an $N = 1$ supergravity framework,
interpreted as a low-energy effective field theory. In minimal
supergravity models, the K{\"a}hler function $K$ that describes the
kinetic terms for the chiral supermultiplets $\Phi \equiv (\zeta, \phi)$,
where the $\zeta$ represent hidden-sector fields and the $\phi^i$
observable-sector fields, has the form $K = \Sigma_i |\Phi^i|^2$. We 
denote derivatives of $K$ with respect to the chiral superfields by $K_i 
\equiv \partial K / \partial \phi^i$, etc. In the minimal supergravity
case, we have $K^i = {\phi^i}^* + {W^i}/W$, $K_i = \phi_i + W_i^*/W^*$,
and $({K^{-1}})^j_i = \delta ^j_i$, and the resulting scalar potential is 
(in units where the Planck mass is unity)
\beq
V(\phi,\phi^*) \; = \; e^{ \phi_i {\phi^i}^*} \left[
|W^i + {\phi^i}^* W |^2 - 3|W|^2 \right].
\label{msgpot}
\eeq
It is then apparent that the soft supersymmetry-breaking
scalar masses $m_0$ are universal at the input GUT scale, with~\cite{BIM}
\beq
m_0^2 \; = \; m_{3/2}^2,
\label{msugra}
\eeq
where $m_{3/2}$ is the gravitino mass and we assume that the tree-level
cosmological constant vanishes. If we further assume that the
superpotential $W(\Phi)$ may be separated into pieces $f$ and $g$ that are
functions only of observable-sector fields $\phi^i$ and hidden-sector
fields $\zeta$, respectively, then the soft supersymmetry-breaking
trilinear terms $A_0$ and bilinear terms $B_0$ are also universal, and are
related by~\cite{BIM}
\beq
B_0 \; = \; A_0 - m_{3/2},
\label{BA}
\eeq
so that
\beq
{\hat B} \; = \; {\hat A} - 1,
\label{BAhat}
\eeq
which is one of the principal options we studied in~\cite{AB1} and 
discuss further below.

The simplest model for local supersymmetry breaking in minimal
supergravity~\cite{pol} has just one additional chiral multiplet $\zeta$
in addition to the observable matter fields $\phi_i$, with a
superpotential that is separable in this so-called Polonyi field and the
observable fields $\phi_i$: $W = f(\phi) + g(\zeta)$. It takes the simple 
form
\beq
g(\zeta) \; = \; \nu(\zeta + \beta),
\label{polonyi}
\eeq
where we impose $\vert \beta \vert = 2 - \sqrt{3}$ to ensure that the
cosmological constant vanishes. Assuming $\beta$ to be positive, and using
$\langle \zeta \rangle = \sqrt{3} - 1$, we have~\cite{bfs} the universal
soft trilinear supersymmetry-breaking terms
\beq
\hat{A} \; = \; (3 - \sqrt{3}) m_{3/2},
\label{PolonyiA}
\eeq
and universal bilinear soft supersymmetry-breaking terms
\beq
\hat{B} \; = \; (2 - \sqrt{3}) m_{3/2},
\label{PolonyiB}
\eeq
whose consequences we explored in~\cite{AB1} and discuss further below.

In the simplest version of the Giudice-Masiero (GM) mechanism \cite{gm}, in addition to 
minimal supergravity kinetic terms in the observable and hidden 
sectors, and a separable superpotential $W = f + g$, one postulates a 
coupling
\beq
K(\phi, \zeta) \; \ni \; \lambda \zeta^\dagger H_1 H_2,
\label{GMcoupling}
\eeq
where $H_{1,2}$ are the two Higgs supermultiplets in the MSSM. Assuming 
that the cosmological constant vanishes, the term (\ref{GMcoupling}) 
generates a Higgs mixing term (\ref{GMlambda}).
This mechanism also yields the 
nonlinear relation between $\hat{B}$ and $\hat{A}$ given in (\ref{GMrelation}),
whose consequences we explore below.

As already remarked, minimal supergravity models predict a relation
(\ref{msugra}) between $m_0$ and the gravitino mass, which is not
necessarily true in the generic CMSSM. This relation enables us to
delineate the regions of VCMSSM parameter space where the LSP is a
neutralino, the lighter ${\tilde \tau}$ or the gravitino. The
astrophysical and cosmological constraints on gravitino dark matter have
been recently re-examined~\cite{GDM,feng}, taking also into account the
constraints on decays of the next-to-lightest supersymmetric particle
(NSP) arising from comparing the BBN and CMB constraints on the
baryon-to-entropy ratio~\cite{CEFO,had}. In our later discussions of
VCMSSMs, we give unified treatments of the parts of $(m_{1/2}, m_0)$
planes where the LSP is a neutralino, the lighter ${\tilde \tau}$ and the
gravitino.

\section{The Electroweak Vacuum in VCMSSMs}

In the general CMSSM, we start with the following set of input parameters
defined at the GUT scale: $m_{1/2}$, $m_0$, $A_0$, $B_0$ and the Higgs
mixing parameter $\mu_0$.  At tree level, the electroweak vacuum is
specified by the following two conditions:
\ba
m_Z^2 & = & {2 (m_1^2 + \mu^2 - (m_2^2 + \mu^2) \tan^2 \beta) \over
(\tan^2 \beta -1)} , \\ 
\sin 2 \beta & = & { - 2 B \mu}/(m_1^2 + m_2^2 + 2 \mu^2) ,
\label{treerel}
\ea
and the pseudoscalar neutral Higgs mass $m_A$ is determined by
\beq
m_A^2 \; = \; m_1^2 + m_2^2 + 2 \mu^2,
\label{mA}
\eeq
where $m_1$ and $m_2$ are the soft supersymmetry-breaking masses for the
two Higgs doublets at the electroweak scale. These as well as $\mu$ and 
$B$ are assumed to be evaluated by renormalization-group equation (RGE) 
running from the input values. One 
may, alternatively, solve for $\mu$ and $B$ in terms of $m_Z$ and $\tan \beta$:
\ba
\mu^2 & = & \frac{m_1^2 - m_2^2 \tan^2 \beta + \frac{1}{2} \mz^2 (1 -
\tan^2 \beta) + \Delta_\mu^{(1)}}{\tan^2 \beta - 1 + \Delta_\mu^{(2)}} 
\nonumber \\
B \mu  & = & -{1 \over 2} (m_1^2  + m_2^2 + 2 \mu^2) \sin 2 \beta + \Delta_B
\label{onelooprel}
\ea
where we have now included the loop corrections $\Delta_B$ and
$\Delta_\mu^{(1,2)}$ required to relate the RGE values to the 
corresponding
quantities evaluated at
$m_Z$~\cite{Barger:1993gh,deBoer:1994he,Carena:2001fw}, and here $m_{1,2}
\equiv m_{1,2}(m_Z)$~\footnote{As observed in~\cite{AB1},
comparisons~\cite{bench} with {\tt ISASUGRA}~\cite{isa} show that our
procedure of minimizing the Higgs potential at the weak scale gives very
similar spectra, also at large $\tan \beta$ and in the focus-point
region.}. In most treatments of the CMSSM, $m_{1/2}$, $m_0$, $A_0$ and
$\tan \beta$ are taken as inputs, and the conditions (\ref{onelooprel})
are used to determine $\mu$, $B$ and the CP-odd Higgs mass $m_A$.

As discussed in~\cite{AB1}, in a VCMSSM where $B$ is determined in advance
in terms of $A$, it is convenient to use the electroweak vacuum conditions
(\ref{onelooprel}) to determine $\tan \beta$ as a function of $m_0$ and
$m_{1/2}$ for some input value of $A$. However, since $\Delta_\mu$ depends
on $\tan \beta$, and $\Delta_B$ depends on both $\mu$ and $\tan \beta$ in 
a
nonlinear way, it is not possible to write down an analytical solution for
$\tan \beta$. Moreover, it was shown in~\cite{dn1} using an RGE-improved
tree-level calculation for $\tan \beta$ in the minimal supergravity model
that there may be up to three possible solutions for $\tan \beta$ for any
given choices of $m_{1/2}, m_0$, and $A_0$. We remarked
previously~\cite{AB1} that we typically find just one solution with a
moderately low value of $\tan \beta$, that multiple solutions exist only
for $m_t < 153$ GeV, and that $B$ always increases monotonically with
$\tan \beta$ over the range $5 < \tan \beta < 55$ in our calculations.
Thus, a given value of $m_{1/2}$, $m_0$, $A_0$ and $sgn(\mu)$ always
corresponds, in our analysis, to a definite value for $\tan \beta$. Since 
this is important for our treatment of VCMSSMs, we now illustrate this 
point in more detail.

\begin{figure}
\begin{center}
\mbox{\epsfig{file=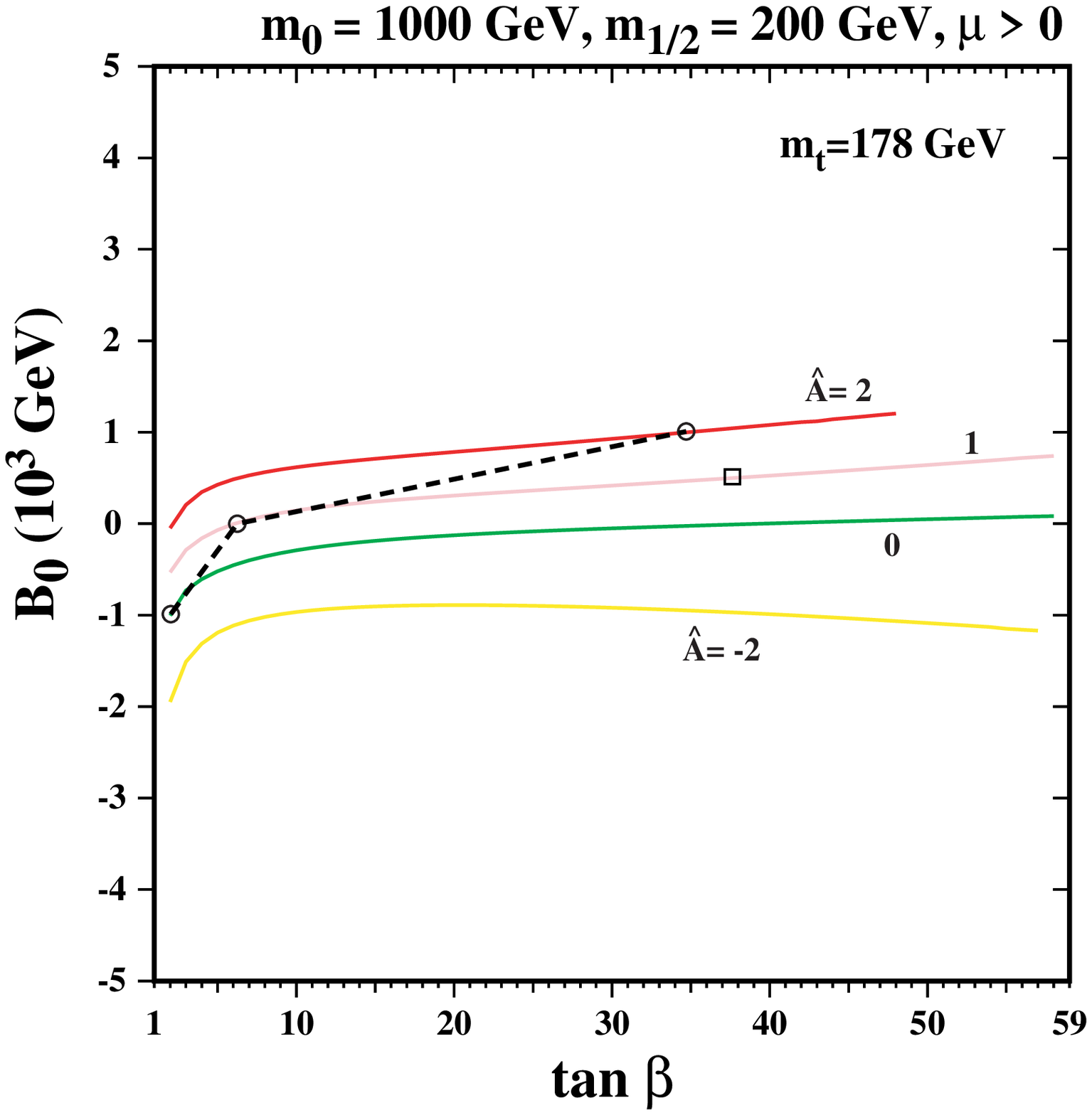,height=7.5cm}}
\mbox{\epsfig{file=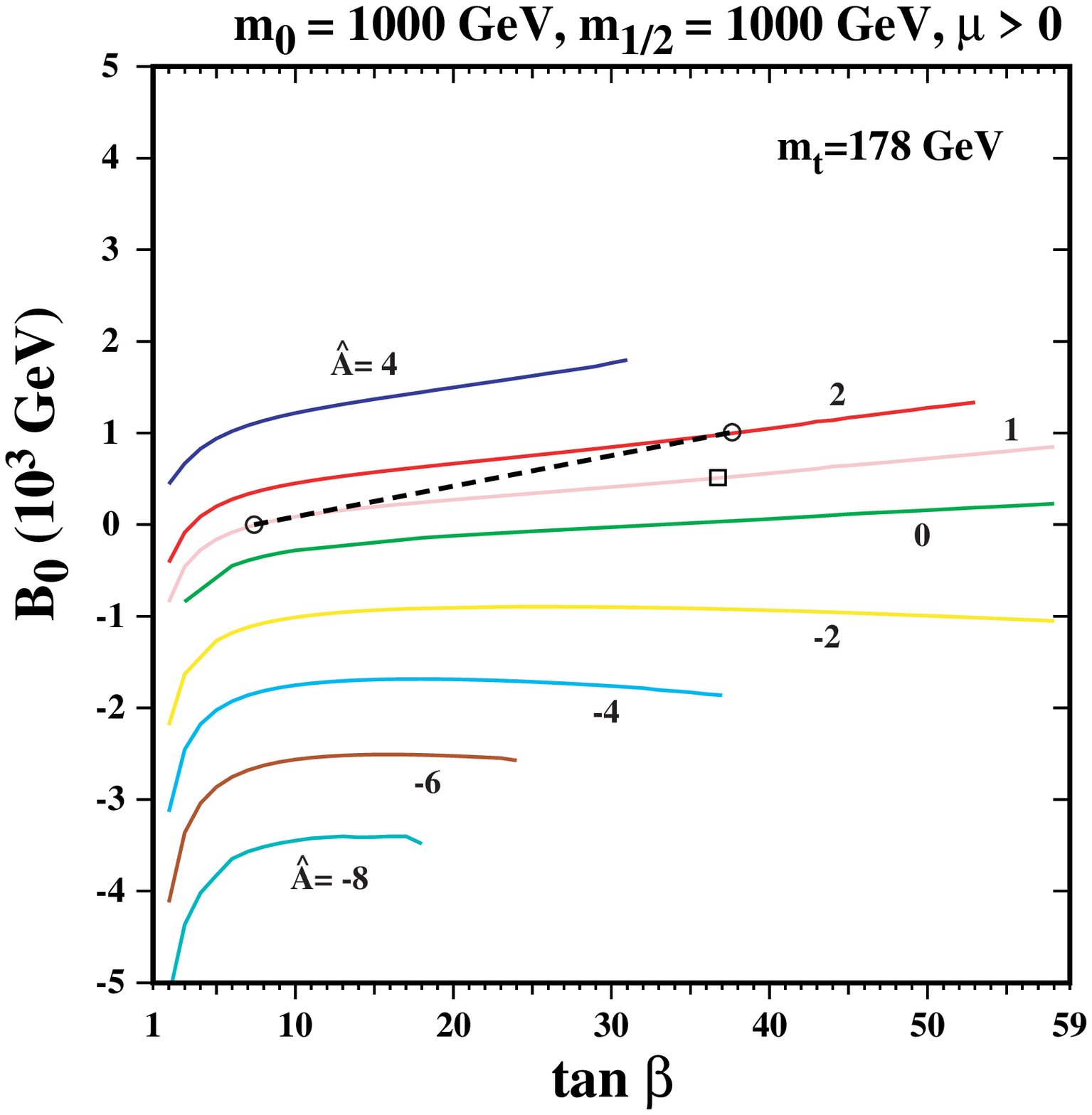,height=7.5cm}}
\end{center}
\begin{center}
\mbox{\epsfig{file=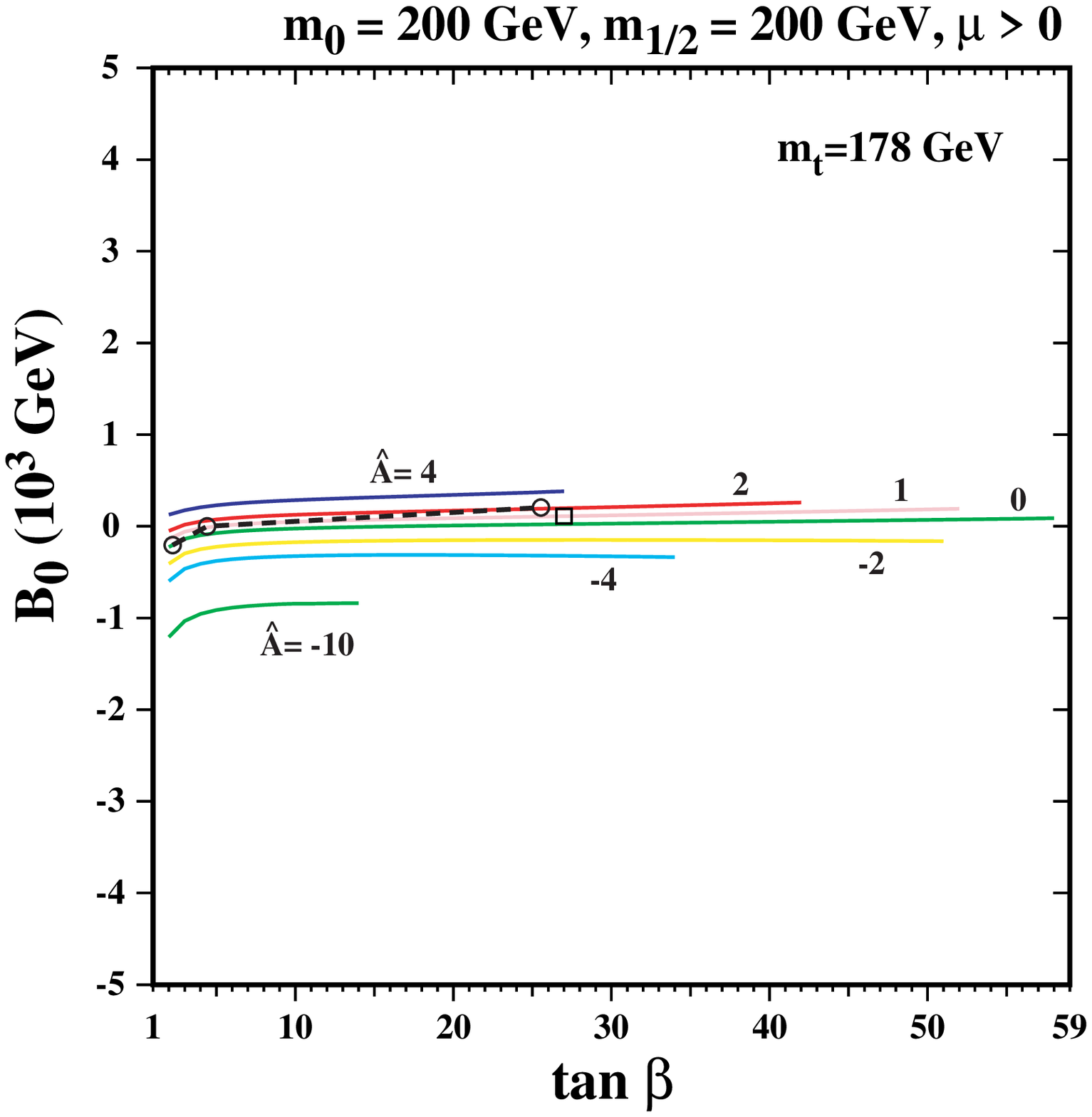,height=7.5cm}}
\mbox{\epsfig{file=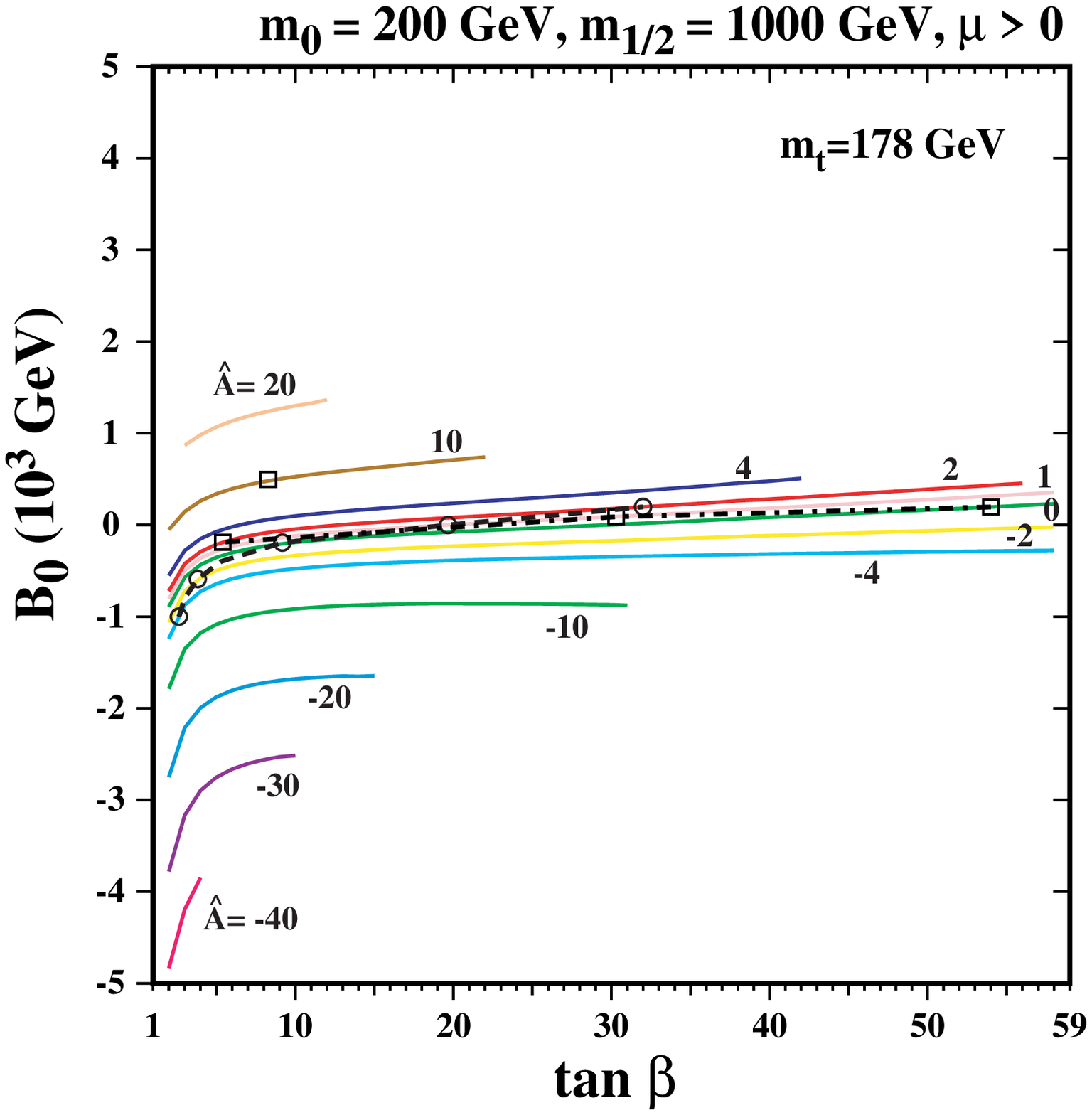,height=7.5cm}}
\end{center}
\caption{\label{fig:Btbmup}\it
Values of $B_0$ as functions of $\tan \beta$ for $\mu >0$ and 
$(m_{1/2}, m_0) =$ (a) $(200, 1000)$~GeV, (b) $(1000, 1000)$~GeV, (c) 
$(200, 200)$~GeV and (d) $(1000, 200)$~GeV. Solutions for ${\hat B} = {\hat A} -
1$ case are denoted by small circles, which are connected by dashed lines. 
Solutions in the case of the Giudice-Masiero mechanism are denoted by 
small squares, connected by dot-dashed lines when possible.}
\end{figure}

\begin{figure}
\begin{center}
\mbox{\epsfig{file=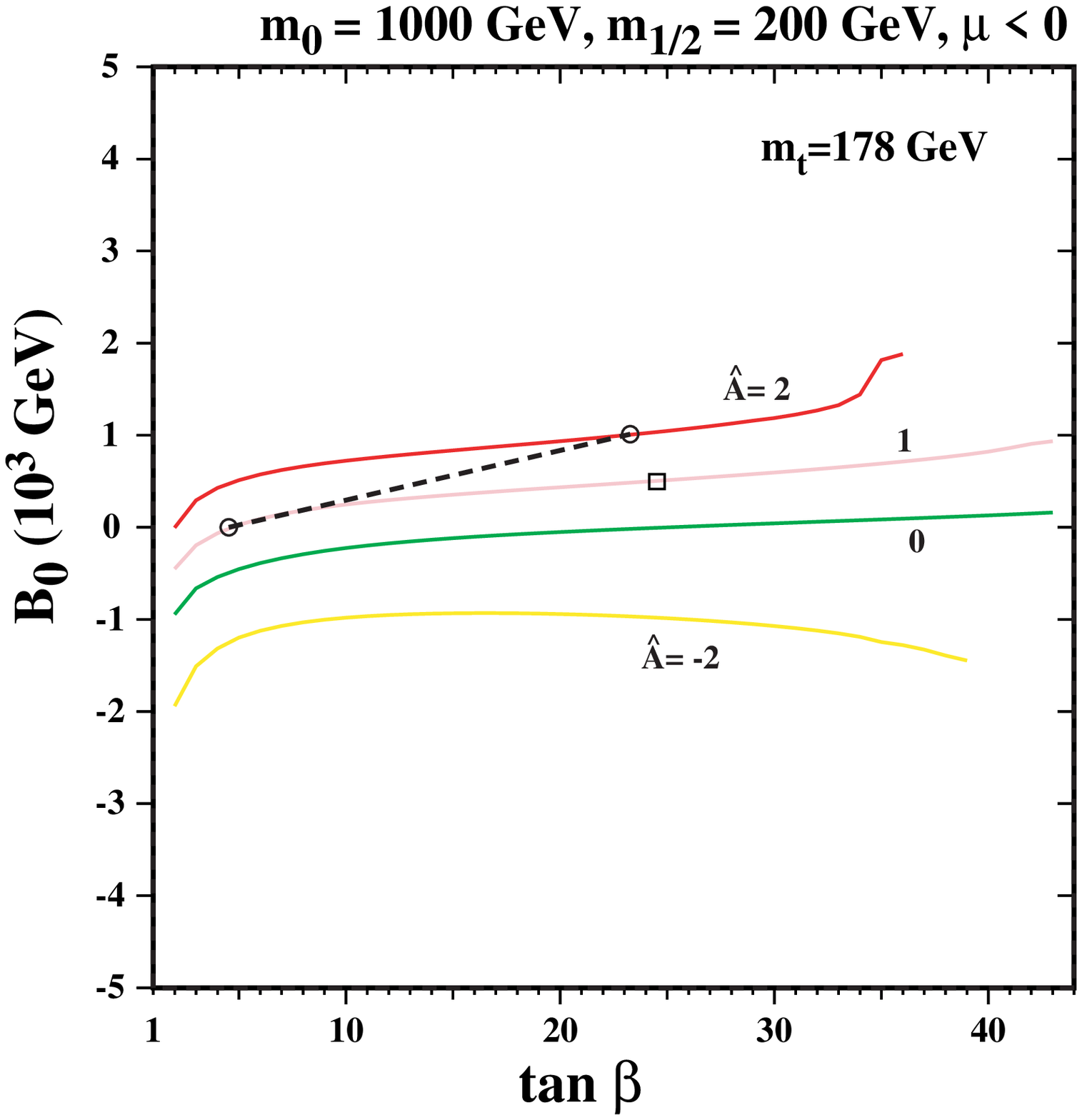,height=7.5cm}}
\mbox{\epsfig{file=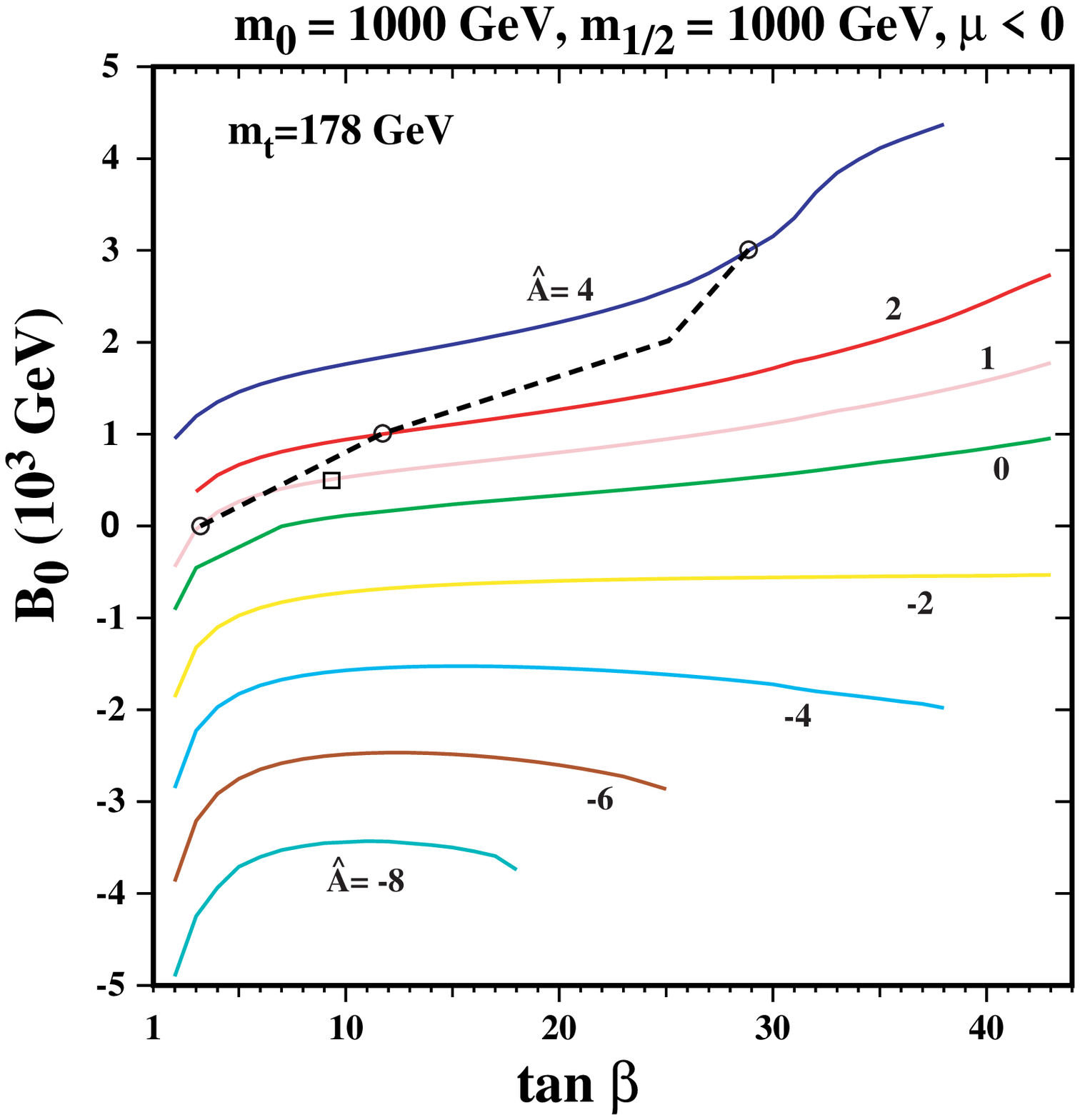,height=7.5cm}}
\end{center}
\begin{center}
\mbox{\epsfig{file=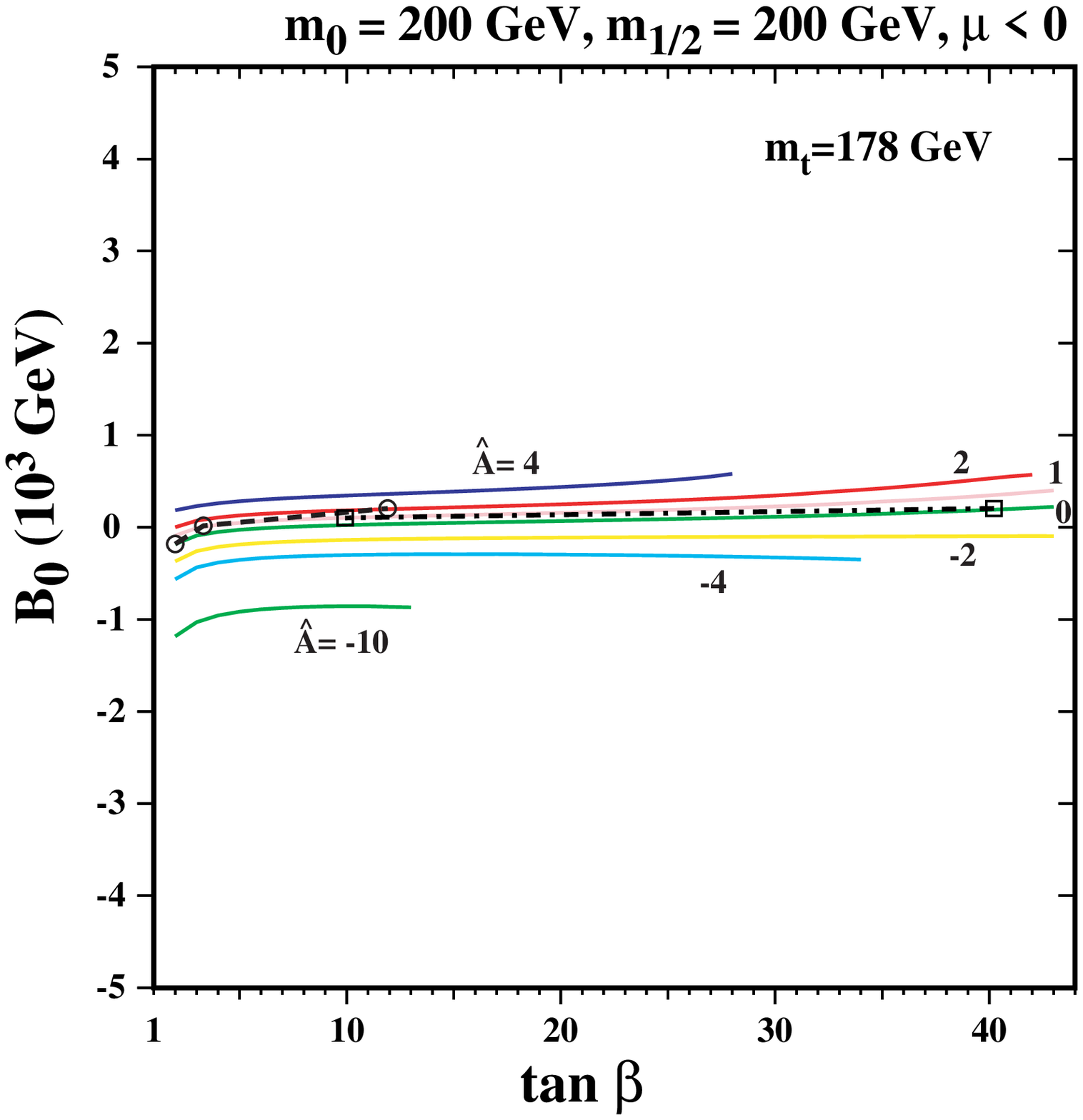,height=7.5cm}}
\mbox{\epsfig{file=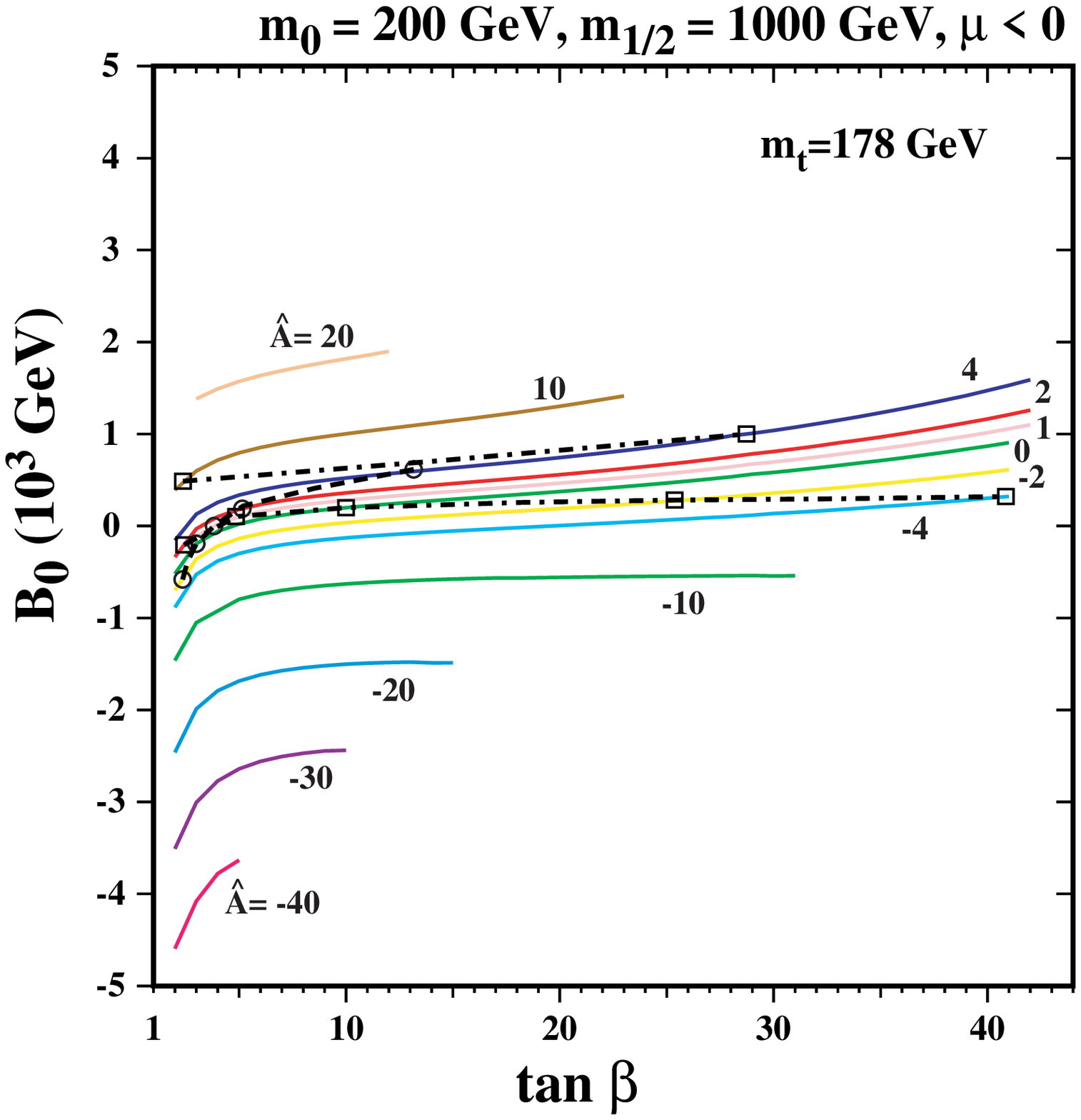,height=7.5cm}}
\end{center}
\caption{\label{fig:Btbmun}\it
As in Fig.~\ref{fig:Btbmup}, but for $\mu < 0$.
}
\end{figure}

We show in Fig.~\ref{fig:Btbmup} some examples of the necessary input
values of $B_0$ as functions of $\tan \beta$, for four representative
choices of $(m_0, m_{1/2})$ and $\mu > 0$. We use $m_t = 178$~GeV as suggested
by the latest CDF and D0 results~\cite{mtop}. We see that $B_0$ generally
increases monotonically for all positive values of $A_0$, and also for
some negative values of $A_0$. This is also true for $\mu < 0$, as seen in
Fig.~\ref{fig:Btbmun}. These observations immediately imply that, in any
VCMSSM that predicts a unique value of $B_0$ for a given value of $A_0$,
there will be (at most) a unique value of $\tan \beta$ where the VCMSSM
relation is obeyed. 

We note, however, that there are some particular negative values of $A_0$
for which the required value of $B_0$, after rising when $\tan \beta$ is
small, decreases slightly at large $\tan \beta$. This raises the
possibility that there might be two allowed values of $\tan \beta$ in some
restricted set of VCMSSMs. One example is when $m_{1/2} = 200$~GeV, $m_0 =
1000$~GeV, $\mu > 0$, ${\hat A} \simeq -2$ and ${\hat B} \simeq -1.5$, as
seen in panel (a) of Fig.~\ref{fig:Btbmup}, and there are some other
examples in other panels of Figs.~\ref{fig:Btbmup} and \ref{fig:Btbmun}.
However, in practice, such multiple solutions do not exist in the specific
VCMSSMs that we study in this paper.

To illustrate this more explicitly, we indicate by small circles in
Figs.~\ref{fig:Btbmup} and \ref{fig:Btbmun} the values of $\tan \beta$
where the minimal supergravity condition ${\hat B} = {\hat A} -1$ is
satisfied, for a few specific values of ${\hat A}$, and we join these
points by dashed lines. For each value of ${\hat A}$ there is clearly only
one consistent choice of $\tan \beta$ for any given choice of $(m_{1/2},
m_0)$. We also show solutions for the Giudice-Masiero mechanism case, indicated
by small squares. 

Note that we do not obtain solutions for $\hat B$ for all choices of $\hat A$.
For example,  in Fig.~\ref{fig:Btbmup}a, we show solutions only for $\hat 
A = 0, 1$ and 2 for $\hat B = \hat A - 1$ and $\hat A = 1$ for the 
Giudice-Masiero model. In the minimal
supergravity case, when $\hat A$ is reduced, $\hat B$ is also reduced driving the solution
to smaller values of $\tan \beta$.  Very quickly these solutions drop below
$\tan \beta = 2$ and, below $\tan \beta \sim 1.7$, the RGEs do not provide
solutions to the sparticle spectra due to a divergence in the top quark Yukawa
coupling at the unification scale. Similarly when $\hat A$ is large, the solution is driven to very large
values of $\tan \beta$ where again no solutions to the RGEs are found.
In the case of the GM model, the slope of $B_0$ vs $\tan \beta$ is very small,
and small changes in $\hat A$ lead to large changes in $\tan \beta$. 
Note also that in the GM model,  there are often two branches of solutions
which are disconnected.  This is seen for example in Figs.~\ref{fig:Btbmup}d and \ref{fig:Btbmun}d.
This is due to the relation (\ref{GMrelation}) which separates solutions at $\hat A = 3$. 

\begin{figure}
\begin{center}
\mbox{\epsfig{file=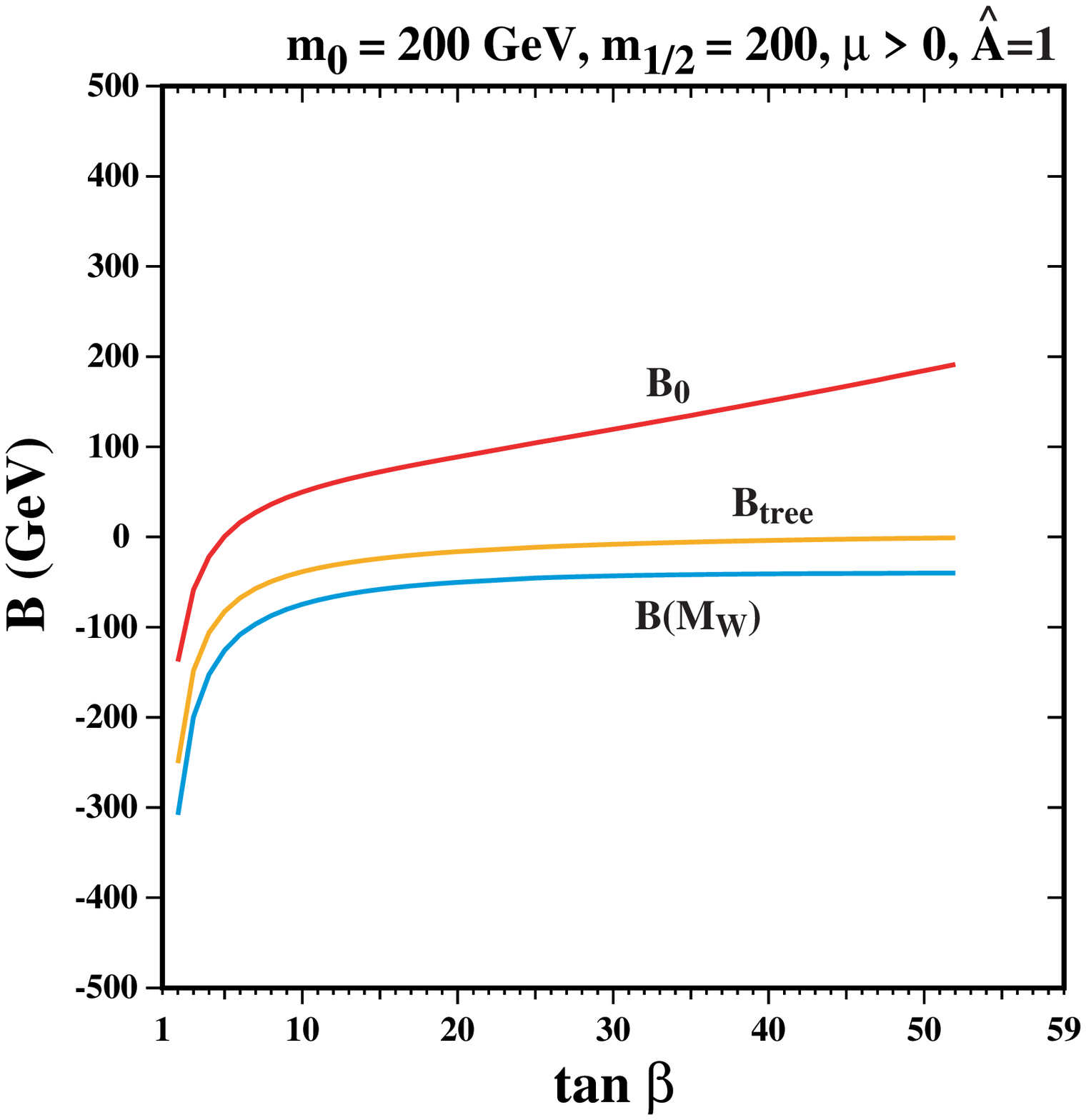,height=7.5cm}}
\mbox{\epsfig{file=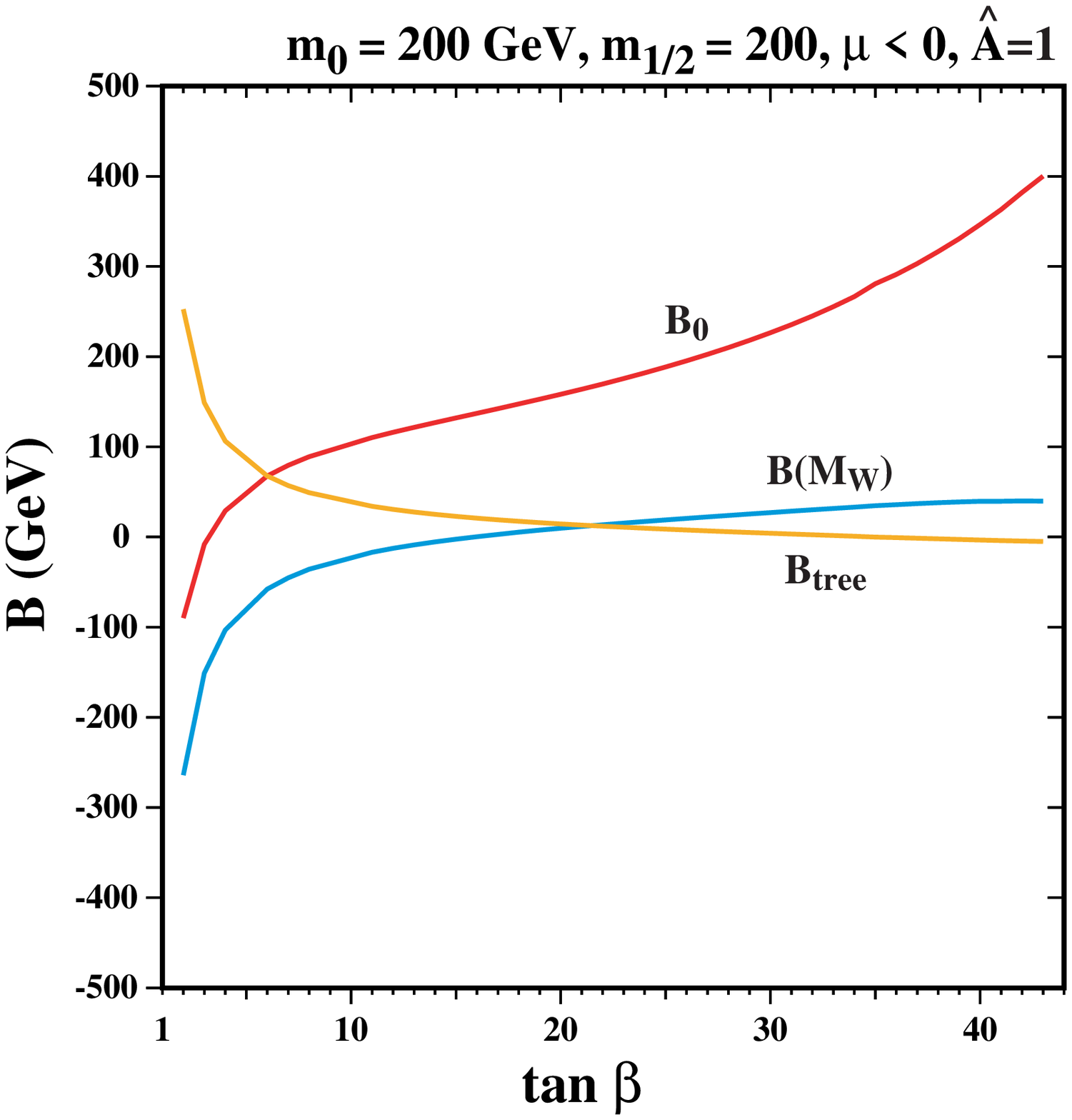,height=7.5cm}}
\end{center}
\caption{\label{fig:BBB}\it
The values of $B_0$, the electroweak tree-level and full $B(M_W)$, as 
functions of $\tan
\beta$ for (a) $\mu > 0$ and (b) $\mu < 0$, both for $(m_{1/2},m_0) =
(200,200)$~GeV and ${\hat A} = 1$. 
}
\end{figure}

In order to have an analytical feel for the solutions for $B_0$ shown in
Fig.~\ref{fig:Btbmup} and \ref{fig:Btbmun}, we show in Fig.~\ref{fig:BBB} 
the values of $B_0$ at the input GUT 
scale, the tree-level values at the electroweak scale
and the full values of $B(M_W)$ as functions of $\tan \beta$, (a) for $\mu 
> 0$ and (b) for $\mu < 0$, in both cases for $(m_{1/2},m_0) = 
(200, 200)$~GeV. The tree-level value of $B$ at the electroweak 
scale is defined as
\beq
B_{tree} \; \equiv \; \frac{(m_1^2 + m_2^2 + 2 \mu^2) \sin 2 \beta}{ - 2 \mu},
\label{Btree}
\eeq
and as one can see $B_{tree}$ tends to 0 as $\tan \beta $ is increased.
The `full' values are calculated including one-loop electroweak 
threshold corrections, and $B_0$ is then the result of running the RGEs 
from the weak scale to the unification scale.
In the $\mu > 0$ case, we see in Fig.~\ref{fig:BBB} that $B_0$ is 
systematically larger than the 
tree-level value of $B(M_W)$, which is in turn larger than its full value. 
However, even in this case $B(M_W)$ increases monotonically with $\tan 
\beta$. The situation is rather different for $\mu < 0$, where we see 
that the sign of the loop correction depends on the value
of $\tan \beta$, vanishing for $\tan \beta \simeq 21$. As a result, 
the full value of $B(M_W)$ and hence $B_0$ increase monotonically with $\tan \beta$.
Had we neglected the 1-loop corrections to $B$ and ran the RGEs up to the unification 
scale, we could obtain a non-monotonic solution for $B_0$ with respect to
$\tan \beta$ (for example, a solution with a minimum value of $B_0$) leading
to multiple solutions of $\tan \beta$ for a fixed value of $\hat A$~\cite{th}.  Thus,
Fig.~\ref{fig:BBB} indicates the importance of the loop correction in
determining the number of solutions. 

\section{Updated Constraints on the CMSSM}

The standard LEP constraints and cosmological constraints on the CMSSM
have been discussed previously in many places~\cite{eoss,wmapothers}, so we do
not discuss them
further here, except to recall that we use the WMAP range $0.094 <
\Omega_\chi h^2 < 0.129$ \cite{wmap} for the relic density of the LSP, assumed to be
the lightest neutralino $\chi$. However, there are three new experimental
developments that we should like to mention. One is the new value $m_t =
178.0 \pm 4.3$~GeV recently reported by the CDF and D0
collaborations~\cite{mtop},
another is the evolution in the possible discrepancy between the
experimental measurement of $g_\mu - 2$ and the value calculated in the
SM, and the other is a recent improved upper limit on the branching ratio
for $B_s \to \mu^+ \mu^-$.

\begin{figure}
\begin{center}
\mbox{\epsfig{file=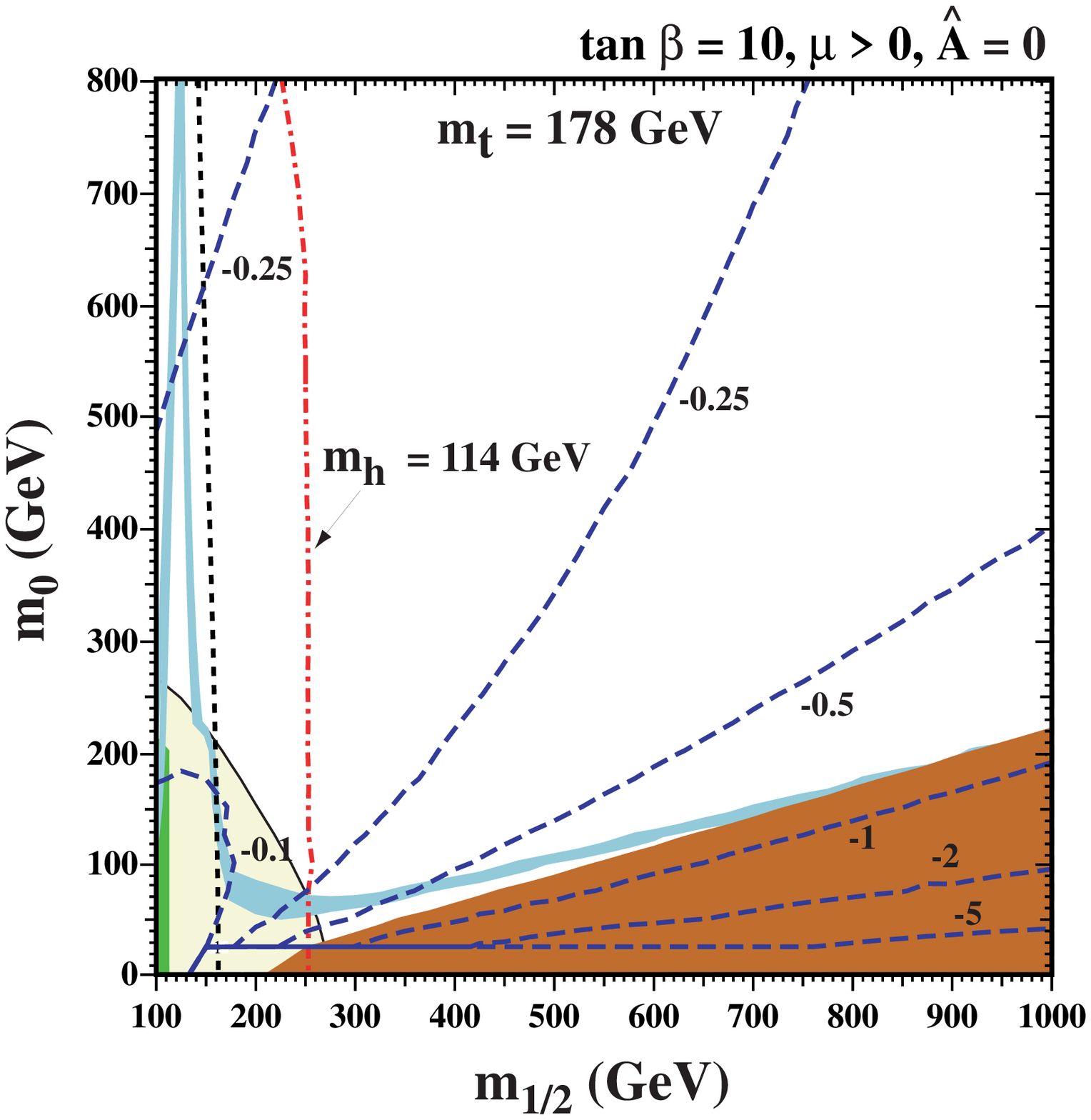,height=8cm}}
\mbox{\epsfig{file=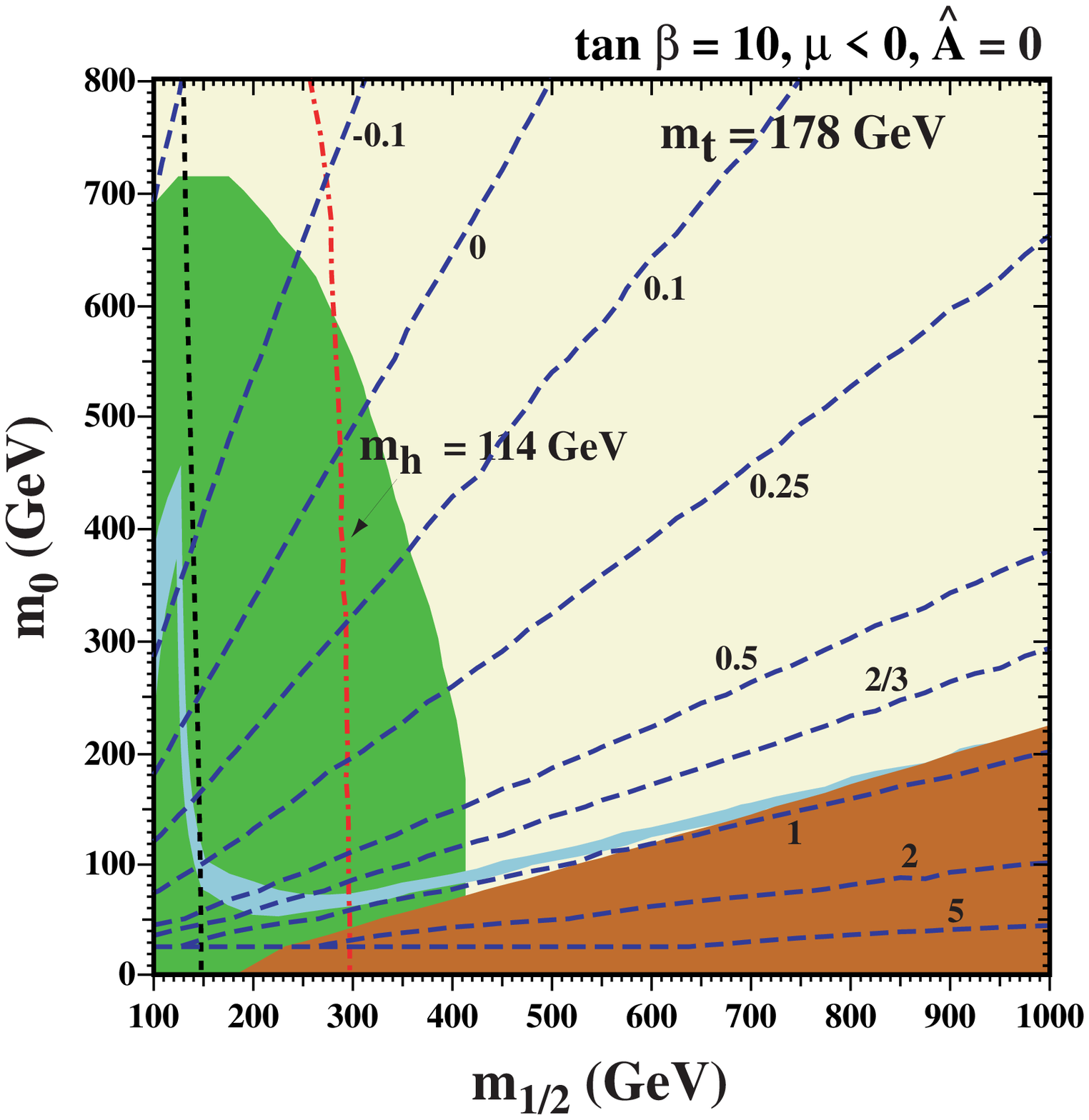,height=8cm}}
\end{center}
\begin{center}
\mbox{\epsfig{file=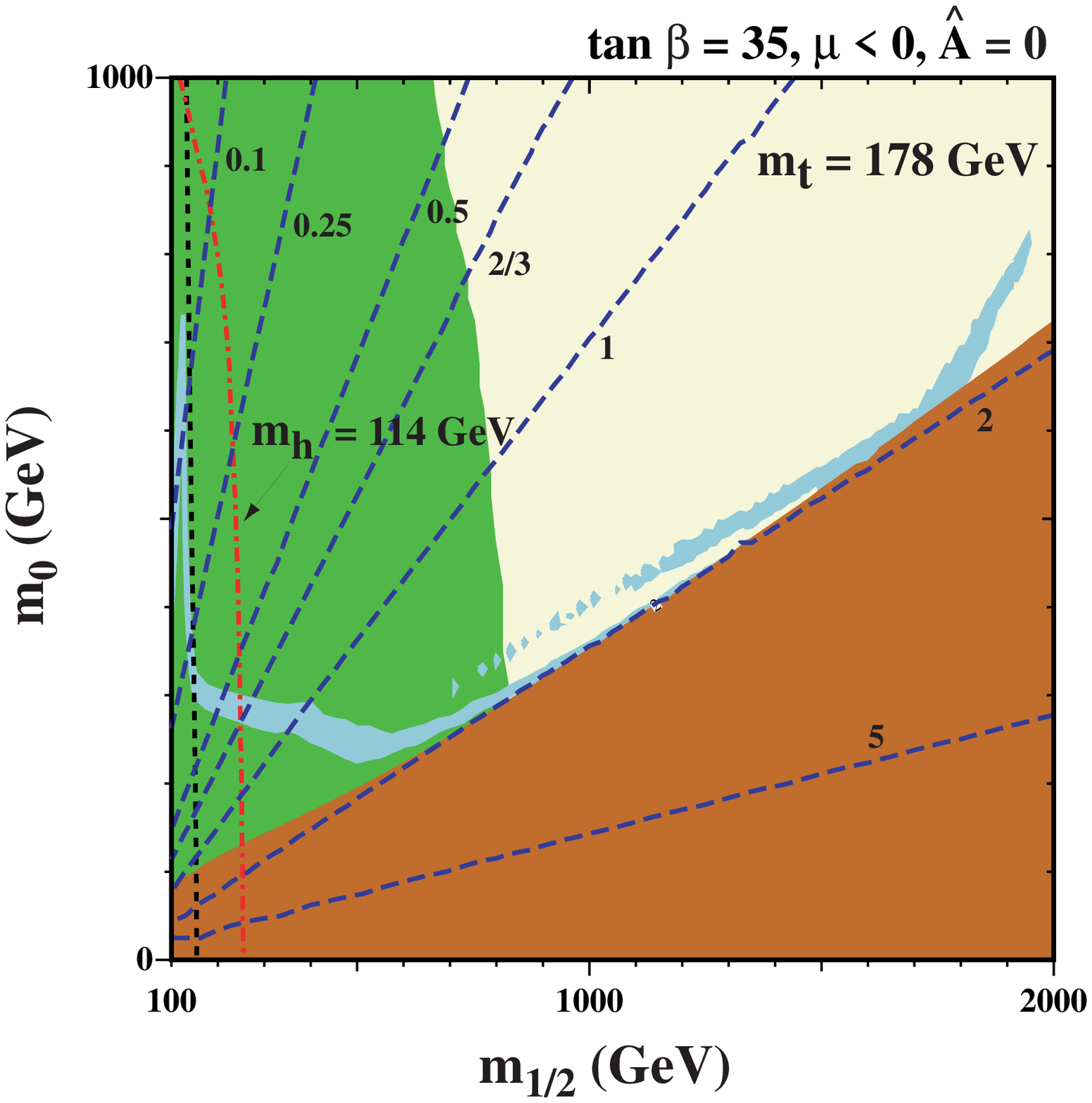,height=8cm}}
\mbox{\epsfig{file=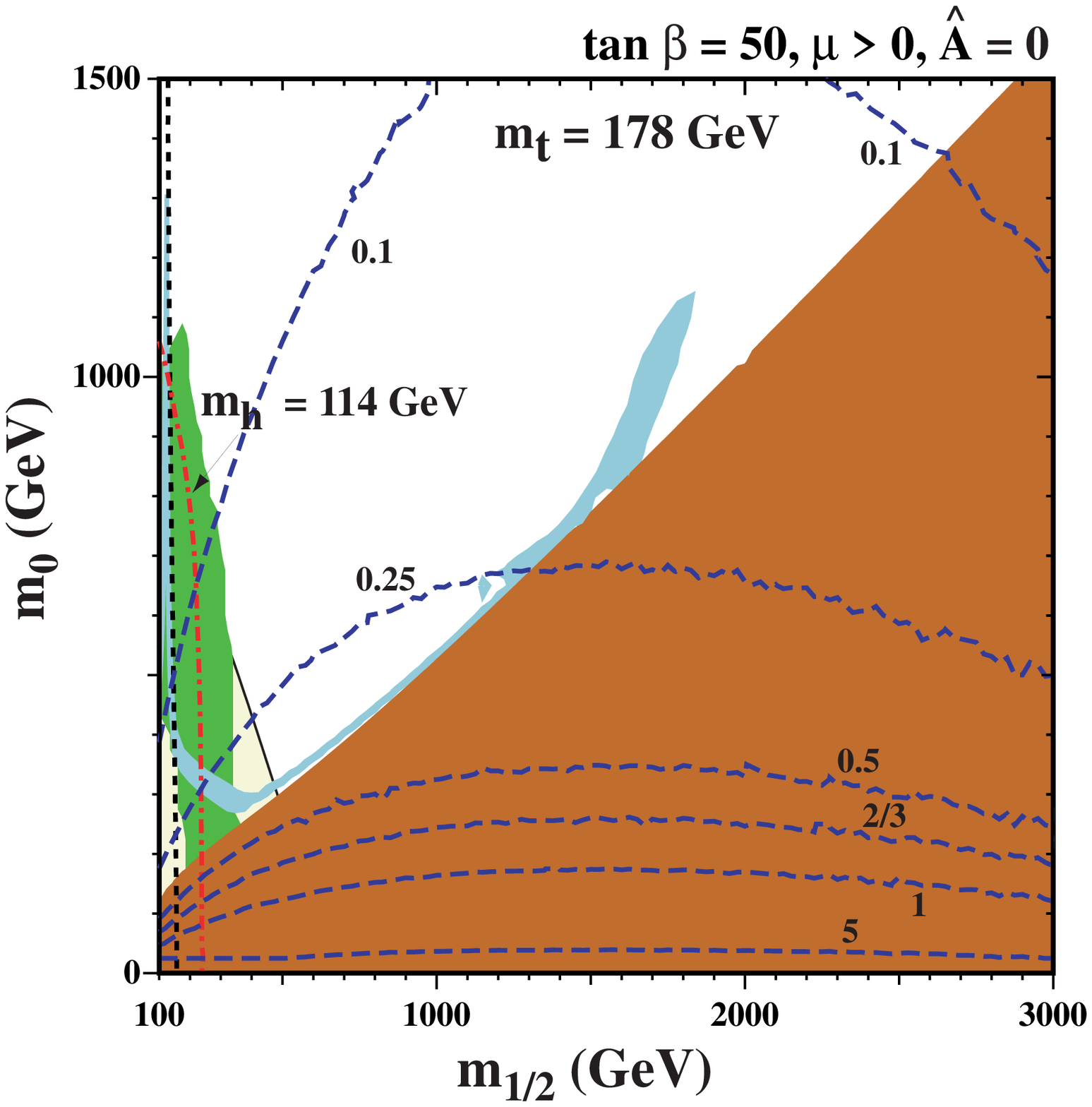,height=8cm}}
\end{center}
\caption{\label{fig:CMSSM}\it
The $(m_{1/2}, m_0)$ planes in the CMSSM for (a) $\tan\beta = 10, \mu > 0$, (b)
$\tan\beta = 10, \mu < 0$, (c) $\tan\beta = 35, \mu < 0$, and (d)
$\tan\beta = 50, \mu > 0$, all for $\hat{A} = 0$. In each panel, the
region allowed by the WMAP cosmological constraint $0.094 \le \Omega_\chi 
h^2 \le 0.129$ has pale (turquoise) shading. The disallowed region where 
$m_{\tilde \tau_1} < m_\chi$ has dark (red) shading.
The regions excluded by $b \rightarrow s \gamma$ have medium
(green) shading, and those in panels (a,d) that are disfavoured by $g_\mu 
- 2$ at the 2-$\sigma$ 
level have very light (yellow) shading with a thin (black) border. 
The contours $m_{\chi^\pm} = 104$~GeV ($m_h = 114$~GeV) are shown
as near-vertical black dashed (red dot-dashed) lines.
In addition, we show several contours of $\hat{B}$ as (blue) 
dashed lines. There is no allowed point compatible with the minimal 
supergravity condition ${\hat B} = {\hat A} - 1$ or the Giudice-Masiero 
model in these plots.}
\end{figure}

The new value of $m_t$ affects the CMSSM parameter space in three
important ways. One is to alter the calculation of the lightest MSSM Higgs
boson mass $m_h$, and hence the lower limit on $m_{1/2}$ inferred from the
LEP lower limit $m_h > 114.4$~GeV. For example, in panel (a) of
Fig.~\ref{fig:CMSSM} for $\tan \beta = 10$ and $\mu > 0$, the lower limit
on $m_{1/2}$ is reduced by about 50~GeV when one increases $m_t$ from
175~GeV to the value of 178~GeV shown here. A second effect is to alter the 
calculation of the rapid-annihilation funnels shown in panels (c) and (d) 
of Fig.~\ref{fig:CMSSM} for $\tan \beta = 35$ and $\mu < 0$ and for 
$\tan \beta = 50$ and $\mu > 0$, respectively. The sensitivity of these
regions to $m_t$ and large $\tan \beta$ was discussed earlier \cite{sven}
in the context of the observability of the Higgs boson at hadron colliders.
Finally, the larger value 
of $m_t$ increases significantly the value of $m_0$ where the focus-point 
region may be found \cite{fp}. For example, for $\tan \beta = 10$ and $\mu > 0$, we 
now find a focus-point region only for $m_0 \gappeq 7$~TeV for $m_{1/2} 
\gappeq 250$~GeV. We do not discuss focus points further in this paper.

The BNL $g_\mu - 2$ experiment recently announced a new determination 
using $\mu^-$ and a final combined value using all their $\mu^\pm$ 
data~\cite{newg2Co}. 
Comparing with the SM calculations of Davier {\it et al.}~\cite{davier}, 
they quote a 
discrepancy of $a_\mu \equiv (g_\mu - 2)/2$ with the SM amounting to
\begin{eqnarray}
\delta a_\mu & = & (27 \pm 10) \times 10^{-10} \; (e^+ e^- {\rm ~data}) 
\nonumber \\
& = & (12 \pm 9) \times 10^{-10} \; (\tau {\rm ~data}).
\label{BNL}
\end{eqnarray}
Another calculation of the SM contribution to $(g_\mu - 2)$ using just the 
$e^+ e^-$ data~\cite{Hag04} yielded a slightly larger discrepancy:
\begin{eqnarray}
\delta a_\mu & = & (32 \pm 10) \times 10^{-10} \; (e^+ e^- {\rm ~data})
\label{Hag}
\end{eqnarray}
There has subsequently been a new SM calculation of the hadronic vacuum 
polarization contribution by de Troc\'oniz and Yndur\'ain~\cite{deTY}, 
who 
quote
\begin{eqnarray}
\delta a_\mu & = & (27 \pm 8) \times 10^{-10} \; (e^+ e^- {\rm ~data}) 
\nonumber \\
& = & (19 \pm 8) \times 10^{-10} \; (\tau {\rm ~and~} e^+ e^- {\rm 
~data}).
\label{TY}
\end{eqnarray}
However, neither of these evaluations include the recent
re-evaluation of the light-by-light contribution to $a_\mu$ by Melnikov 
and Vainshtein~\cite{mv}, which decreases the discrepancy with the SM by about 
$4 \times 10^{-10}$ compared with (\ref{TY}). Therefore, for the purposes of 
the subsequent discussion, we show contours corresponding to
\beq
\delta a_\mu \; = \; (15 \pm 8) \times 10^{-10}.
\label{final}
\eeq
We exhibit this constraint at the 2-$\sigma$ level, in which case its
effect is essentially to exclude the option $\mu < 0$ but allow most of the
$(m_{1/2}, m_0)$ plane for $\mu > 0$, apart from a region of small
$m_{1/2}$ and $m_0$. However, we are well aware that the range
(\ref{final}) is open to question, particularly in view of the discrepancy
between the estimates of the SM contribution based on $e^+ e^-$ and $\tau$
data, and, to a lesser extent, the uncertainty in the light-by-light
contribution. Therefore, we use (\ref{final})  only as an indication, and
by no means a rigid constraint on the parameter space of the CMSSM or any
VCMSSM. In particular, we do not discard the option $\mu < 0$.

Finally, we note that the CDF Collaboration have recently published an
improved experimental upper limit on the branching ratio for $B_s \to
\mu^+ \mu^-$~\cite{bmumu}, namely $5.8 \times 10^{-7}$. Since the branching
ratio
for this decay $\propto \tan^6 \beta$ in the CMSSM, this constraint is
potentially important at large $\tan \beta$. We find that this constraint
is currently still `covered' by the constraints from $b \to s \gamma$,
$m_h$ and $g_\mu - 2$, but this situation may change in the near future.

In preparing the $(m_{1/2}, m_0)$ planes in Fig.~\ref{fig:CMSSM} and the
subsequent figures, we have updated our code by making improvements that
have impacts principally in the rapid-annihilation funnels and focus-point
regions~\footnote{Specifically, we now include the full one-loop
corrections to $m_b$ and $m_t$ instead of approximate
expressions~\cite{pbmz}, and we correct a minor coding error.}. Their
effects are smaller than the other effects mentioned above.

We show in Fig.~\ref{fig:CMSSM} the $(m_{1/2}, m_0)$ planes for a popular
set of CMSSM cases, namely (a) $\tan\beta = 10, \mu > 0$, (b) $\tan\beta =
10, \mu < 0$, (c) $\tan\beta = 35, \mu < 0$, and (d) $\tan\beta = 50, \mu
> 0$, all for $\hat{A} = 0$~\footnote{We note in panels (c) and (d) the 
appearance of allowed bands above the $\chi - {\tilde \tau}_1$ 
coannihilation 
strips, which are due to rapid ${\tilde {\bar\tau}}_1 {\tilde \tau}_1 \to 
H$ 
annihilation.}. In each panel, as well as the `standard'
experimental and cosmological constraints, we have indicated some
representative contours of $\hat{B}$ by (blue) dashed lines. We see that
${\hat B} \gappeq 0$ in almost all the $(m_{1/2}, m_0)$ planes exhibited,
which is clearly incompatible with the minimal supergravity condition
${\hat B} = {\hat A} - 1$. This exemplifies the point that parameter 
choices allowed in the `standard'
CMSSM are often not allowed in favoured VCMSSMs. {\it Specifically, 
the
CMSSM cases shown in Fig.~\ref{fig:CMSSM} could not be realized in minimal
supergravity}: one needs to choose smaller values of $\tan \beta$. A 
similar
conclusion applies to the Giudice-Masiero model, 
which would require ${\hat B} = 1$ for the case ${\hat A} = 0$ considered 
here, 
although GM solutions are possible for $\mu < 0$ if one discards the $g_\mu - 2$ 
constraint and chooses $\tan \beta$ somewhat above 10.

\begin{figure}
\begin{center}
\mbox{\epsfig{file=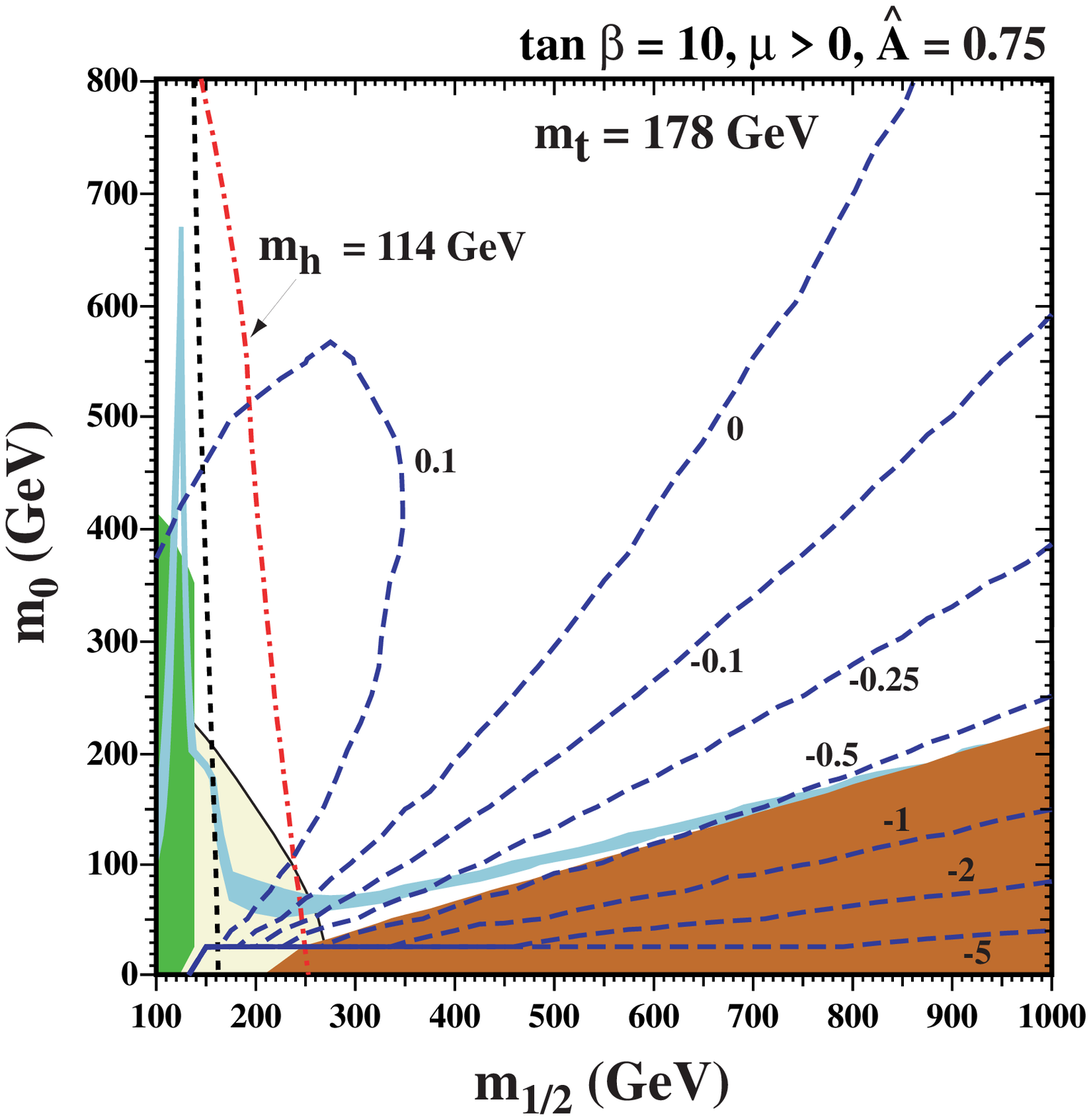,height=8cm}}
\mbox{\epsfig{file=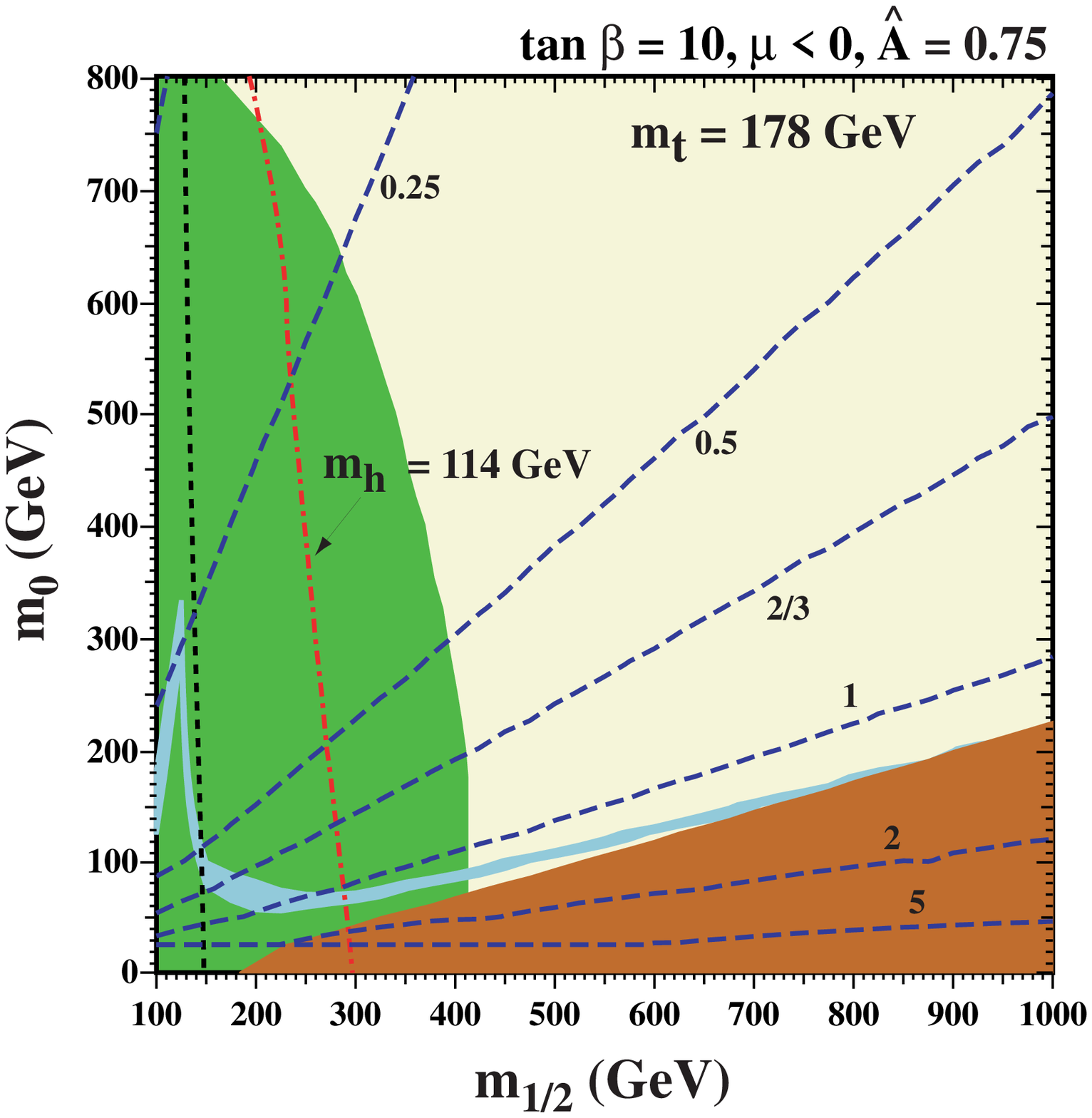,height=8cm}}
\end{center}
\begin{center}
\mbox{\epsfig{file=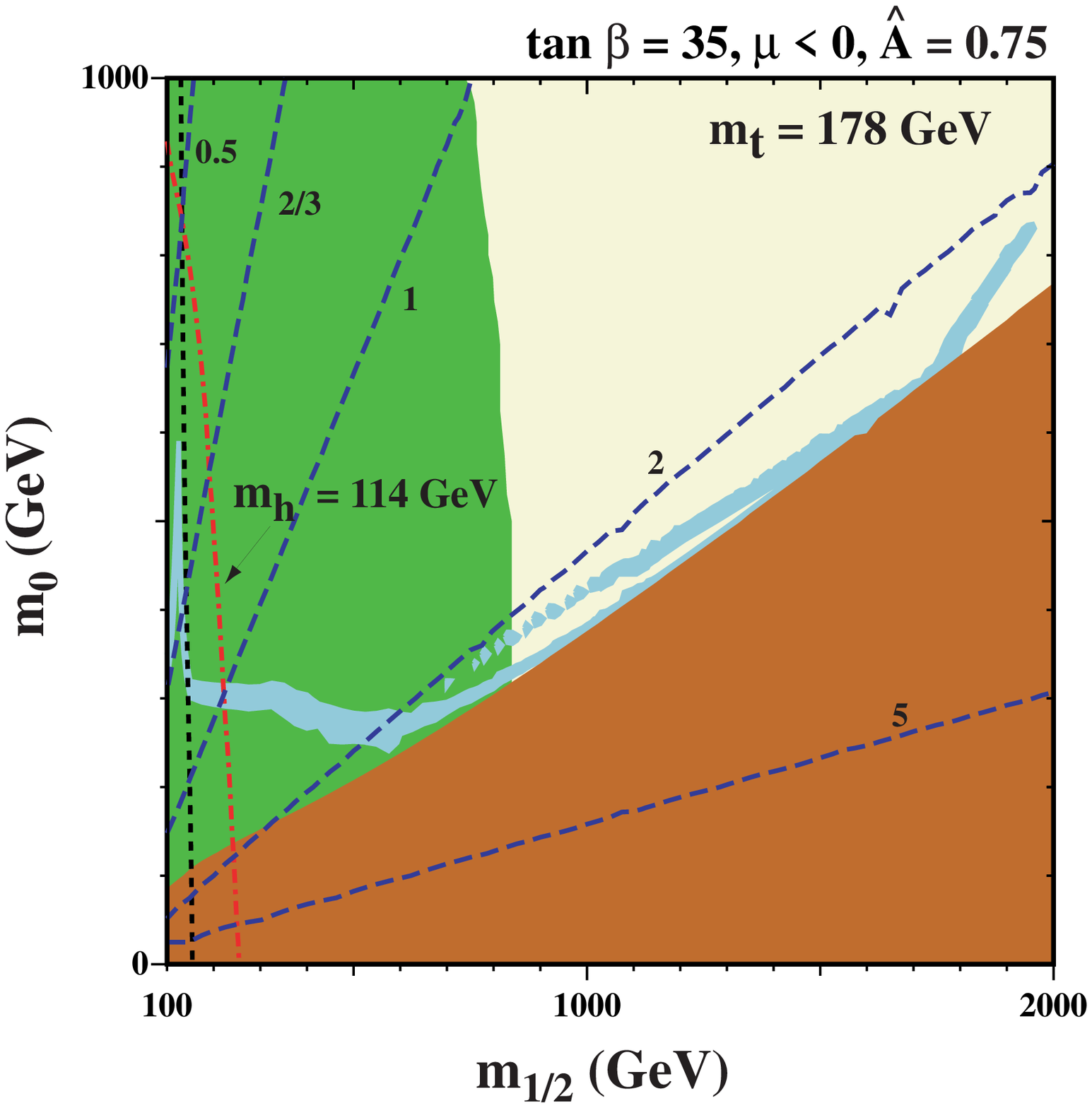,height=8cm}}
\mbox{\epsfig{file=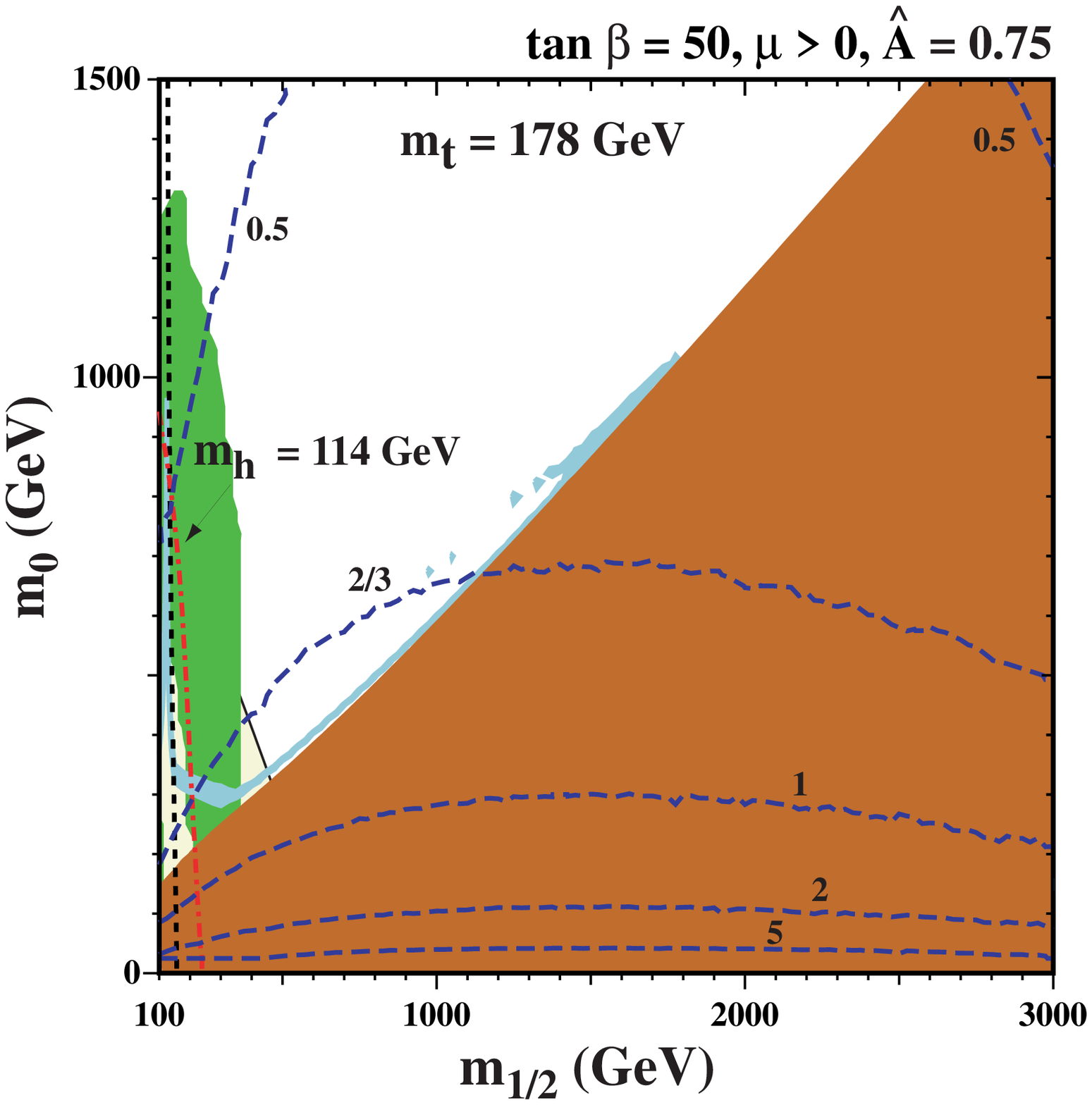,height=8cm}}
\end{center}
\caption{\label{fig:CMSSM2}\it
The $(m_{1/2}, m_0)$ planes in the CMSSM for the same values of 
$\tan\beta$ and the sign of $\mu$ as in Fig.~\ref{fig:CMSSM}, but here for 
$\hat{A} = + 0.75$. }
\end{figure}

Fig.~\ref{fig:CMSSM2} shows the corresponding $(m_{1/2}, m_0)$ planes for 
the same choices of $\tan\beta$ and the sign of $\mu$, but for $\hat{A}
= + 0.75$. In this case, the laboratory and cosmological constraints are 
not greatly different, even in the rapid-annihilation funnel 
regions~\footnote{We note again the rapid ${\tilde \tau}_1 {\tilde {\bar 
\tau}}_1 \to H$ 
annihilation strips in panels (c) and (d).}. 
Now, however, there are some points where the minimal supergravity 
condition ${\hat B} = {\hat A} - 1$ is satisfied, as shown by the 
intersection of the $\hat B = -0.25$ line with the WMAP coannihilation 
region in Fig.~\ref{fig:CMSSM2}a. For this choice of ${\hat A}$, 
the Giudice-Masiero model is also satisfied at a limited number of 
points, exemplified by the intersection of the $\hat B = 2/3$ line with 
the WMAP funnel region in Fig.~\ref{fig:CMSSM2}d.

\section{Examples of $(m_{1/2}, m_0)$ Planes in VCMSSMs}

We now discuss the impacts of the above constraints on some specific
VCMSSMs within the general framework of minimal supergravity, in which
${\hat B} = {\hat A} - 1$. As usual, we display these constraints in
$(m_{1/2}, m_0)$ planes. For the reasons discussed earlier, we regard
$\tan \beta$ as a dependent quantity that varies across these planes,
rather than being a fixed quantity as in most CMSSM analyses. Another
difference from most CMSSM analyses is that the latter generally consider only
the possibility that the LSP is the lightest neutralino $\chi$, assuming
implicitly that the gravitino mass $m_{3/2}$ is sufficiently large that
the gravitino LSP possibility can be neglected. However, in minimal
supergravity, one has $m_{3/2} = m_0$ (\ref{msugra}) if the cosmological 
constant 
$\Lambda = 0$, and the identity of the LSP varies over  the $(m_{1/2}, 
m_0)$ plane. We have
recently published an analysis which includes the possibility that the gravitino
is the LSP possibility~\cite{GDM}, taking
into account the constraints imposed by Big-Bang nucleosynthesis (BBN)  
and the cosmic microwave background (CMB) data on decays of the
next-to-lightest sparticle (NSP) into the gravitino, as well as the relic
gravitino dark matter density itself. In this paper, we incorporate this
analysis into a unified treatment of the neutralino and gravitino LSP
regions of the $(m_{1/2}, m_0)$ planes in VCMSSMs.

\begin{figure}
\begin{center}
\mbox{\epsfig{file=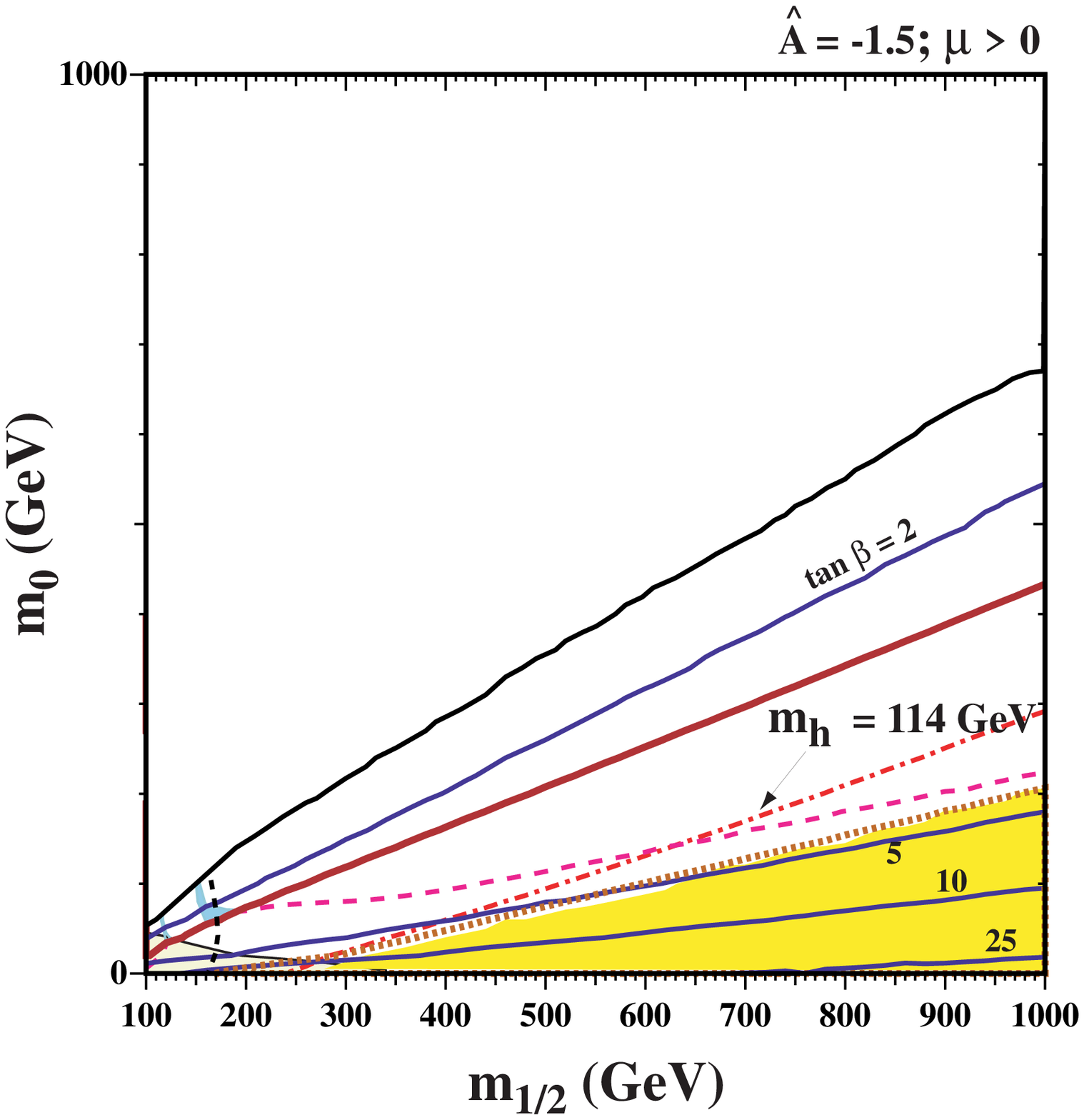,height=8cm}}
\mbox{\epsfig{file=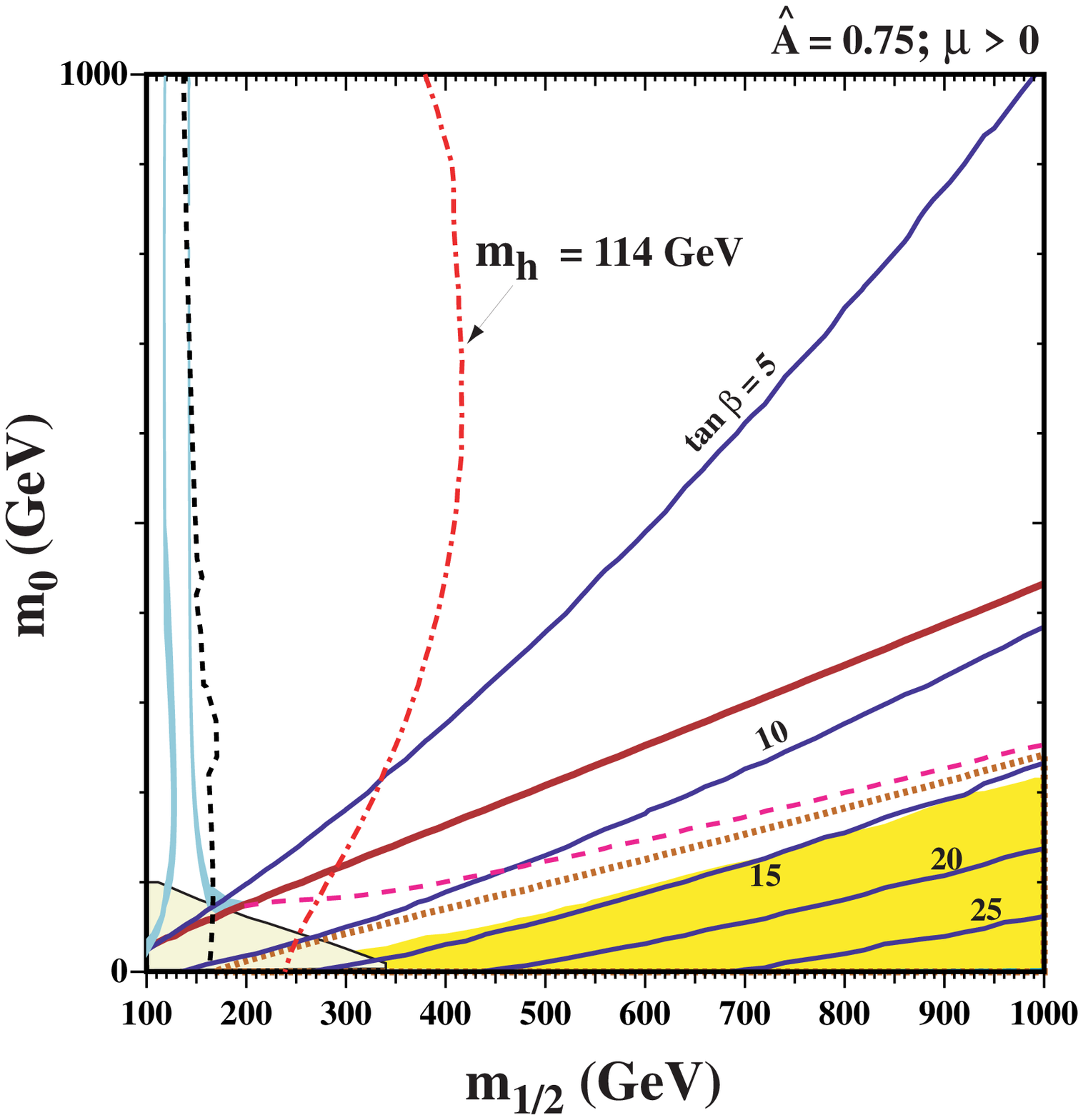,height=8cm}}
\end{center}
\begin{center}
\mbox{\epsfig{file=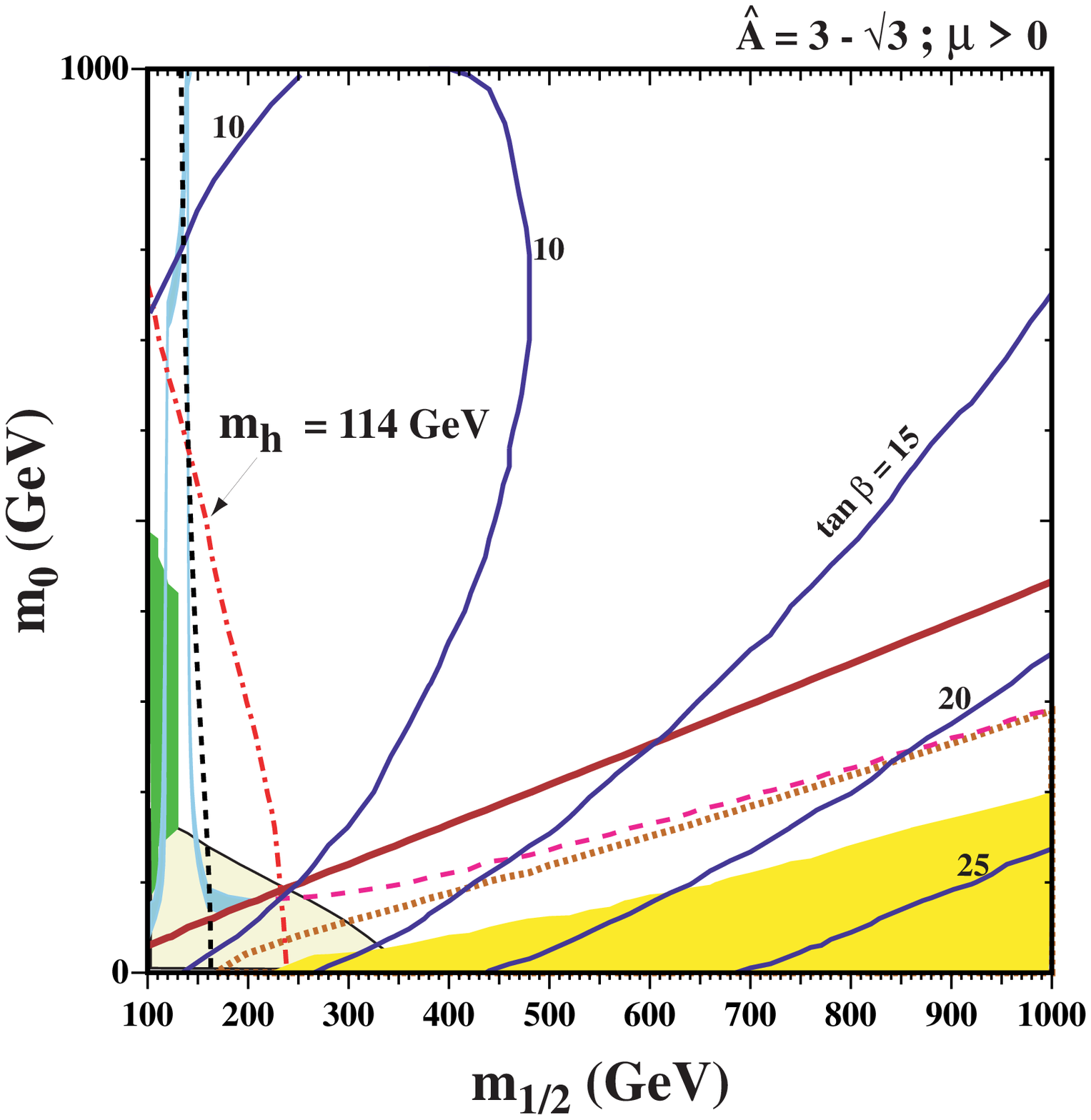,height=8cm}}
\mbox{\epsfig{file=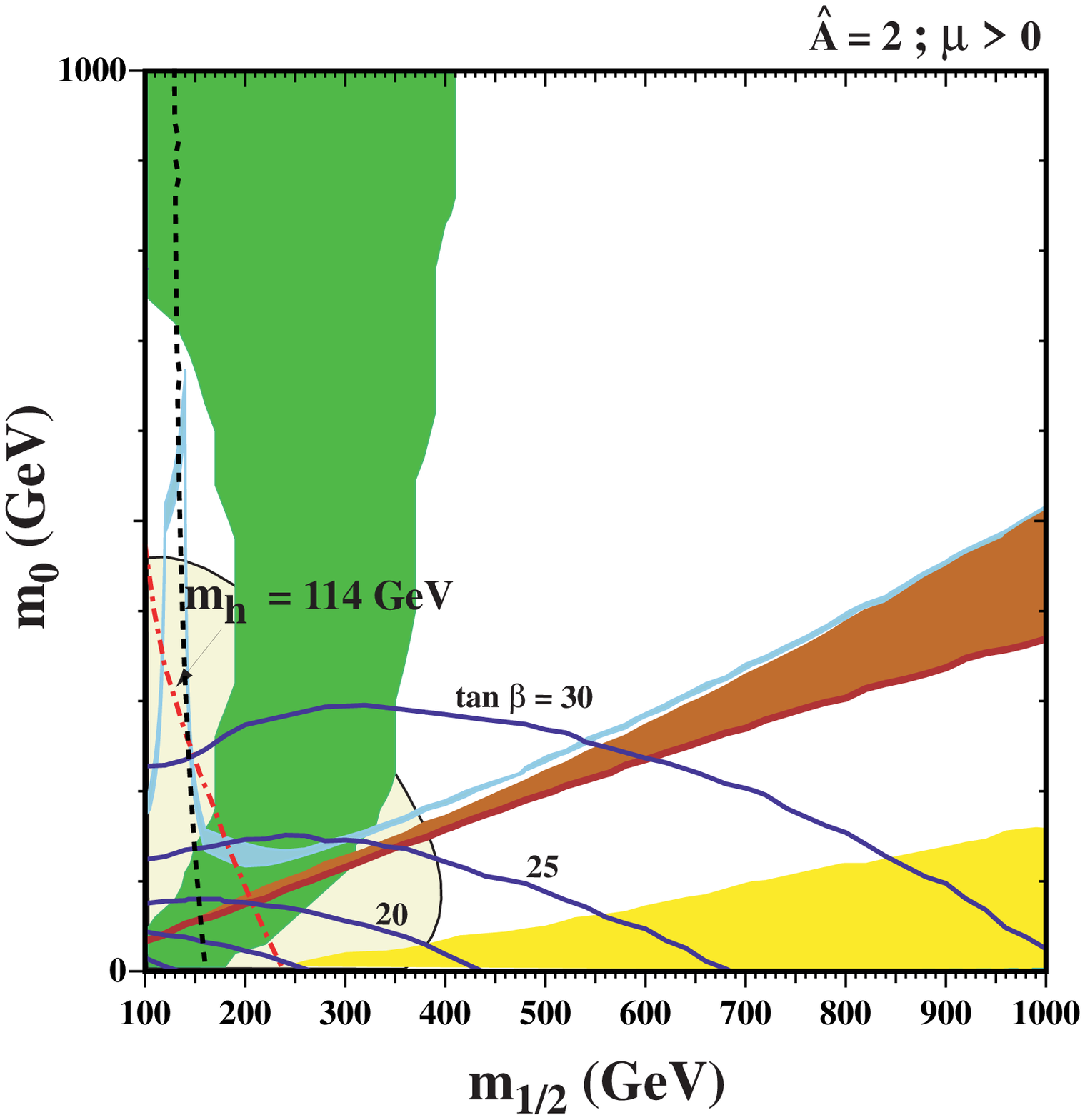,height=8cm}}
\end{center}
\caption{\it
Examples of VCMSSM $(m_{1/2}, m_0)$ planes with contours of $\tan \beta$ 
superposed, for $\mu > 0$ and (a) ${\hat A} = - 1.5$, (b) ${\hat A} = 0.75$, 
(c) the simplest Polonyi model with ${\hat A} = 3 - 
\sqrt{3}$,  and (d) ${\hat A} = 2.0$, all with $ {\hat B} =
{\hat A} -1$. In each panel, we show the regions excluded by 
the LEP lower limits on MSSM particles and those ruled out by $b
\to s \gamma$ decay~\protect\cite{bsg} (medium green shading): the regions 
disfavoured by $g_\mu - 2$ are very light (yellow) shaded, bordered by a thin
(black) line.
The dark (chocolate) solid lines in panels (a, b, c) separate the 
gravitino LSP regions (below). Panel (d) exhibits a dark (red) wedge 
where the LSP is the rapid ${\tilde \tau}_1$. The regions favoured 
by WMAP in the neutralino LSP case have light (turquoise)
shading. The dashed (pink) line corresponds to the maximum relic density 
for the gravitino LSP, and regions allowed by BBN/CMB constraint on NSP 
decay are light (yellow) shaded.}
\label{fig:Polonyi}
\end{figure}

We display in Fig.~\ref{fig:Polonyi} the contours of $\tan \beta$ (solid
blue lines) in the $(m_{1/2}, m_0)$ planes for selected values of ${\hat
A}$, ${\hat B}$ and the sign of $\mu$. Also shown are the contours where
$m_{\chi^\pm} > 104$~GeV (near-vertical black dashed lines) and $m_h >
114$~GeV (diagonal red dash-dotted lines). The regions excluded by $b \to
s \gamma$ have medium (green) shading, and those where the relic density
of neutralinos lies within the WMAP range $0.094 \le \Omega_\chi h^2 \le
0.129$ have light (turquoise) shading. 
The
gravitino LSP and the neutralino LSP regions are separated by dark (chocolate)
solid lines, and the WMAP relic-density strip for neutralinos is shown only above
these lines.
The regions disfavoured by $g_\mu -
2$ at the 2-$\sigma$ level are very light (yellow) shaded. 

If ${\hat A}$ has a large negative value, we do not find any consistent
solutions to the electroweak vacuum conditions. This is reflected in panel
(a) of Fig.~\ref{fig:Polonyi}, for $\mu > 0$ and ${\hat A} = -1.5$, 
where there are no solutions above the
topmost solid (black) line. The solid (blue) contours of $\tan \beta$ rise
diagonally from low values of $(m_{1/2}, m_0)$ to higher values, with
higher values of $\tan \beta$ having lower values of $m_0$ for a given
value of $m_{1/2}$. The dash-dotted (red) $m_h = 114$~GeV contour rises in
a similar way, and regions above and to the left of this contour have $m_h
< 114$ GeV and are excluded. In particular, a neutralino LSP is excluded
in this case. We exhibit in this and the other panels a gravitino LSP
region, which was not studied in our previous exploration of
VCMSSMs~\cite{AB1}. The relic density is acceptably low only below the
dashed (pink) line. This excludes a supplementary domain of the $(m_{1/2},
m_0)$ plane, but the strongest constraint is provided by the BBN/CMB decay
constraint (light, yellow shading), which requires $\tan \beta \gappeq 4.5$.
In panels (b, c, d), the $m_h$ contour rises more vertically, but only in
panel (d) is there any allowed neutralino LSP region. Panel (d) features
an excluded dark (red) shaded wedge where the LSP is the ${\tilde
\tau}_1$.

When ${\hat A}$ is increased to 0.75, as seen in panel (b) of
Fig.~\ref{fig:Polonyi}, both the $\tan \beta$ and $m_h$ contours rise more
rapidly with $m_{1/2}$.  Again, there is no allowed neutralino LSP region.
Within the gravitino LSP region, the $m_h$ and relic density constraints
would both be compatible with $\tan \beta \gappeq 7.5$, but the BBN/CMB
decay constraint imposes the stronger constraint that $\tan \beta \gappeq
13$. It is instructive to compare this figure with Fig.~\ref{fig:CMSSM2}a,
which both assume that $\hat A = 0.75$. The most notable difference is
that, here, fixing the gravitino mass to equal $m_0$ excludes the
neutralino coannihilation region with $\hat B = - 0.25$ and allows a
region of the $(m_{1/2}, m_0)$ plane that would previously have been
excluded because the LSP would have been the ${\tilde \tau}_1$.

An analogous pattern is seen in the simplest Polonyi model with ${\hat A}
= 3 - \sqrt{3}$ shown in panel (c) of Fig.~\ref{fig:Polonyi}, where we
note that the $\tan \beta$ contours have noticeable curvature.  Once
again, the neutralino LSP region is excluded, now by a combination of the
Higgs and chargino mass bounds.  At low $m_0$ in the gravitino LSP region,
the $m_h$ and relic gravitino density constraints impose $\tan \beta
\gappeq 10$ and the BBN/CMB decay constraint imposes $\tan \beta \gappeq
14$~\footnote{There is also a negative Polonyi solution with ${\hat A}
= - 3 + \sqrt{3}$, whose $(m_{1/2}, m_0)$ plane is qualitatively similar 
to panel (a) of Fig.~\ref{fig:Polonyi}.}.

We consider finally the case ${\hat A} = 2.0$ shown in panel (d) of
Fig.~\ref{fig:Polonyi}. In this case, there is a neutralino LSP region in
the $\chi - {\tilde \tau}$ coannihilation strip. Without the $g_\mu - 2$
constraint, the most severe
constraint on this region is imposed by $b \to s \gamma$, requiring $\tan
\beta \gappeq 25$, the $m_h$ constraint being much weaker. 
Imposing the $g_\mu - 2$ constraint requires $\tan
\beta \gappeq 27$. 
There is an
excluded dark (red) shaded wedge where the LSP is the ${\tilde \tau}_1$.
Below this appears a gravitino LSP region with acceptable relic density.
Within this region, the $m_h$ and BBN/CMB decay constraints impose $\tan
\beta \gappeq 15$, which would be strengthened to $\tan \beta \gappeq 20$
if one took the $g_\mu - 2$ constraint at face value. This is the shaded region in
the lower right of panel (d).

We find no consistent solutions for values of ${\hat A}$ substantially
greater than 3 (4) when $\mu > 0$ ($\mu < 0$), and negative values of
${\hat A}$ are
not allowed when $\mu < 0$.  These restrictions arise from the behavior of
the relation between $\tan \beta$ and $B_0$ discussed earlier.  Therefore
as $\hat A$ increases, so does the solution for $\tan \beta$ when $\hat B
= \hat A - 1$.  At very large $\tan \beta$, there are no solutions to the
RGEs due to a divergence in the bottom-quark Yukawa coupling.  For small $\hat 
A$ and $\mu < 0$,
the solution is driven to excessively small values of $\tan \beta$, where
again there are no solutions, now due to the divergence in the top Yukawa 
coupling mentioned earlier. The same is true when $\hat A$ is large and negative and
$\mu > 0$, i.e. for $\hat A < -2.5$, $m_0 \lappeq 500$~GeV for $m_{1/2} \leq
1000$~GeV.

\begin{figure}
\begin{center}
\mbox{\epsfig{file=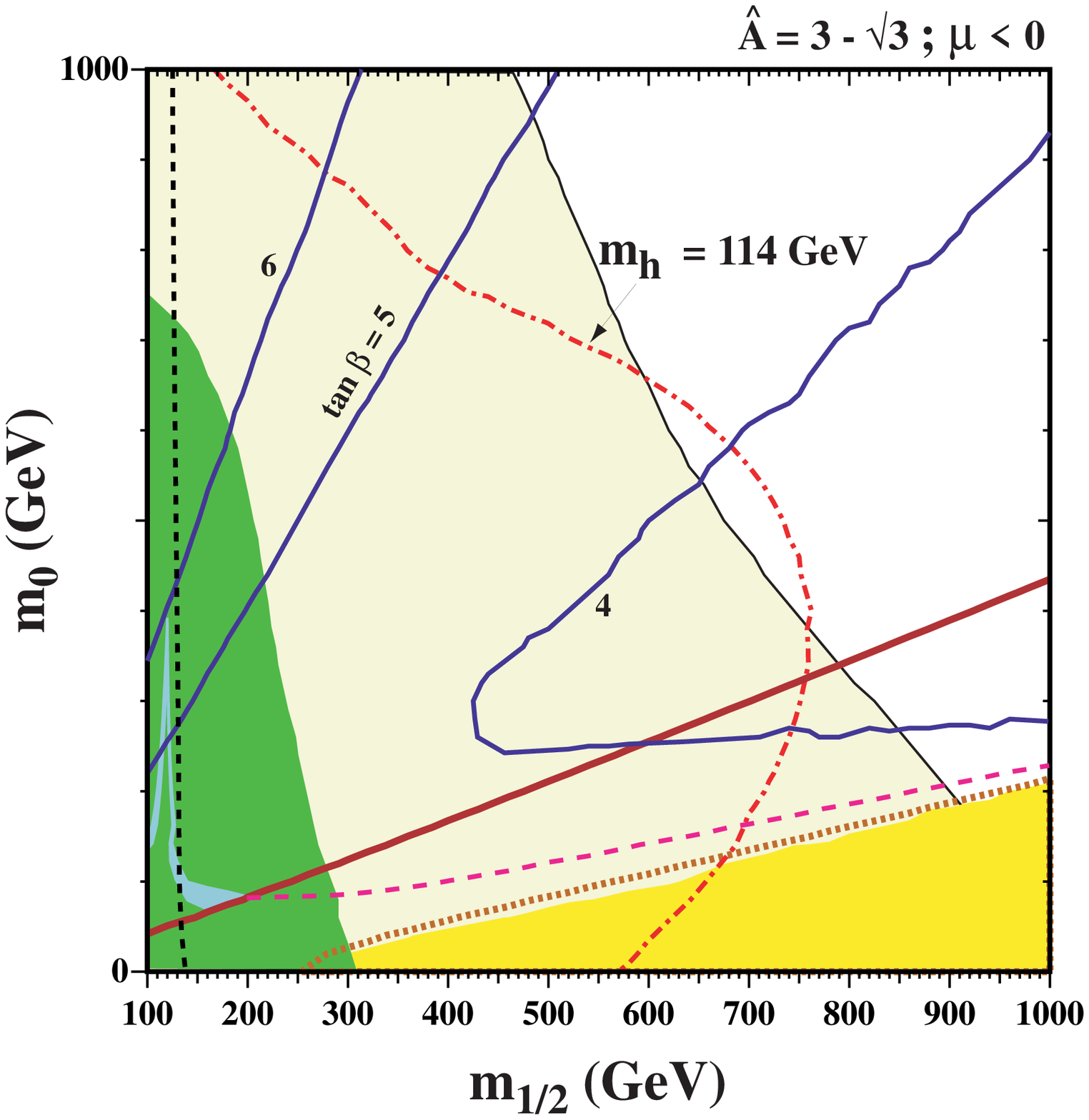,height=8cm}}
\mbox{\epsfig{file=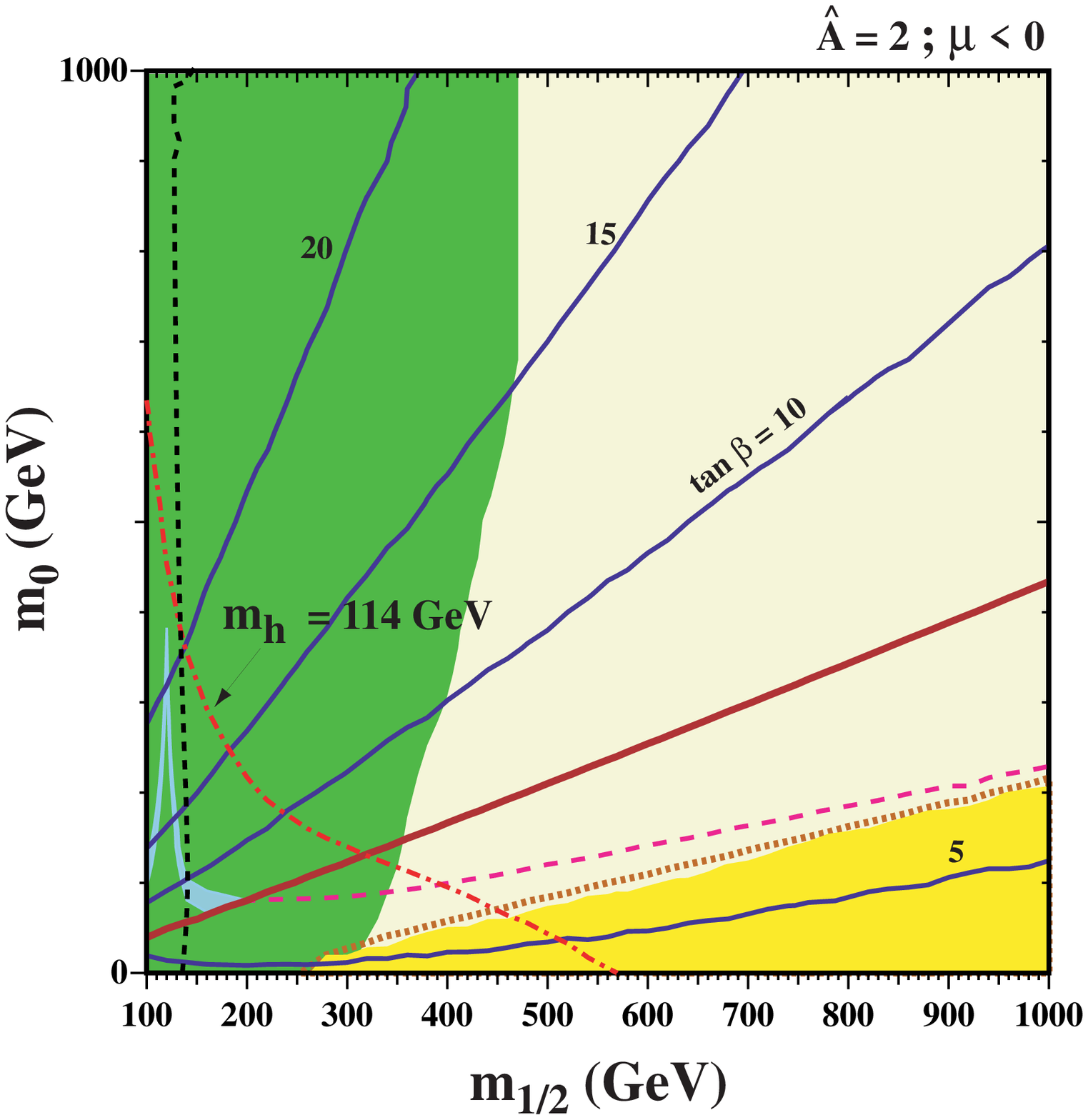,height=8cm}}
\end{center}
\caption{\it
As in Fig.~\ref{fig:Polonyi}, but now for $\mu < 0$ and the
choices (a) ${\hat A} = 3 - \sqrt{3}, {\hat B} =
{\hat A} -1$ and (b) ${\hat A} = 2, {\hat B} =
{\hat A} -1$ and $\mu < 0$.
}
\label{fig:Polonyin}
\end{figure}

As we see in panel (a) of Fig.~\ref{fig:Polonyin}, only a small area
of the $(m_{1/2}, m_0)$ plane in the gravitino LSP region is allowed by
the $m_h$ constraint in the positive Polonyi case ${\hat A} = 3 -
\sqrt{3}$~\footnote{The negative Polonyi case
is not allowed for $\mu < 0$.}.
This area would be further restricted if one took the $g_\mu -
2$ constraint at face value. At larger values of ${\hat A}$, the allowed
region is extended, as exemplified in panel (b) of Fig.~\ref{fig:Polonyin}
for the case ${\hat A} = 2$, where the $m_h$ constraint is somewhat
weaker.  However, in this case the $g_\mu - 2$ constraint would have a
much more drastic effect.

\section{The $\mu$ Problem and the Giudice-Masiero Mechanism}

One of the primary motivations in building supersymmetric model is to
avoid the the gauge hierarchy problem, namely that the Higgs mass is of
order $m_Z$ and much less than the Planck mass, though not protected by 
any symmetry of the Standard Model
between the GUT scale and the weak scale. Supersymmetry alleviates this
problem via cancellations between contributions to the Higgs mass from
fermions and bosons in the same supermultiplet. However, this scenario
begs the question why supersymmetry is broken by soft terms which are
assumed to be $~O(1~{\rm TeV})$. Moreover, there is one other,
supersymmetric, parameter which is required to be small, namely the Higgs
mixing parameter $\mu$. One of the most interesting attempts to explain
the smallness of $\mu$ is the Giudice-Masiero mechanism~\cite{gm}, in
which it is related to a coupling between observable and hidden sectors,
and is of the same order of magnitude as the soft supersymmetry-breaking
parameters. In the simplest realization of the Giudice-Masiero mechanism
with only one hidden superfield, one has the following relation between
${\hat A}$ and ${\hat B}$, as already mentioned:
\beq
\hat{B} = \frac{2 \hat{A} - 3}{\hat{A} - 3},
\eeq
and
\beq
\left| \frac{\mu}{m_0} \right| = \left| \lambda \frac{\hat{A} - 
3}{\sqrt{3}} \right|
\eeq
where $\lambda$ is the coupling constant between the hidden superfield
and the two Higgs supermultiplets. One should require that $\lambda \sim 
O(1)$ for $\mu$ to be the same order of $m_0$.

\begin{figure}
\begin{center}
\mbox{\epsfig{file=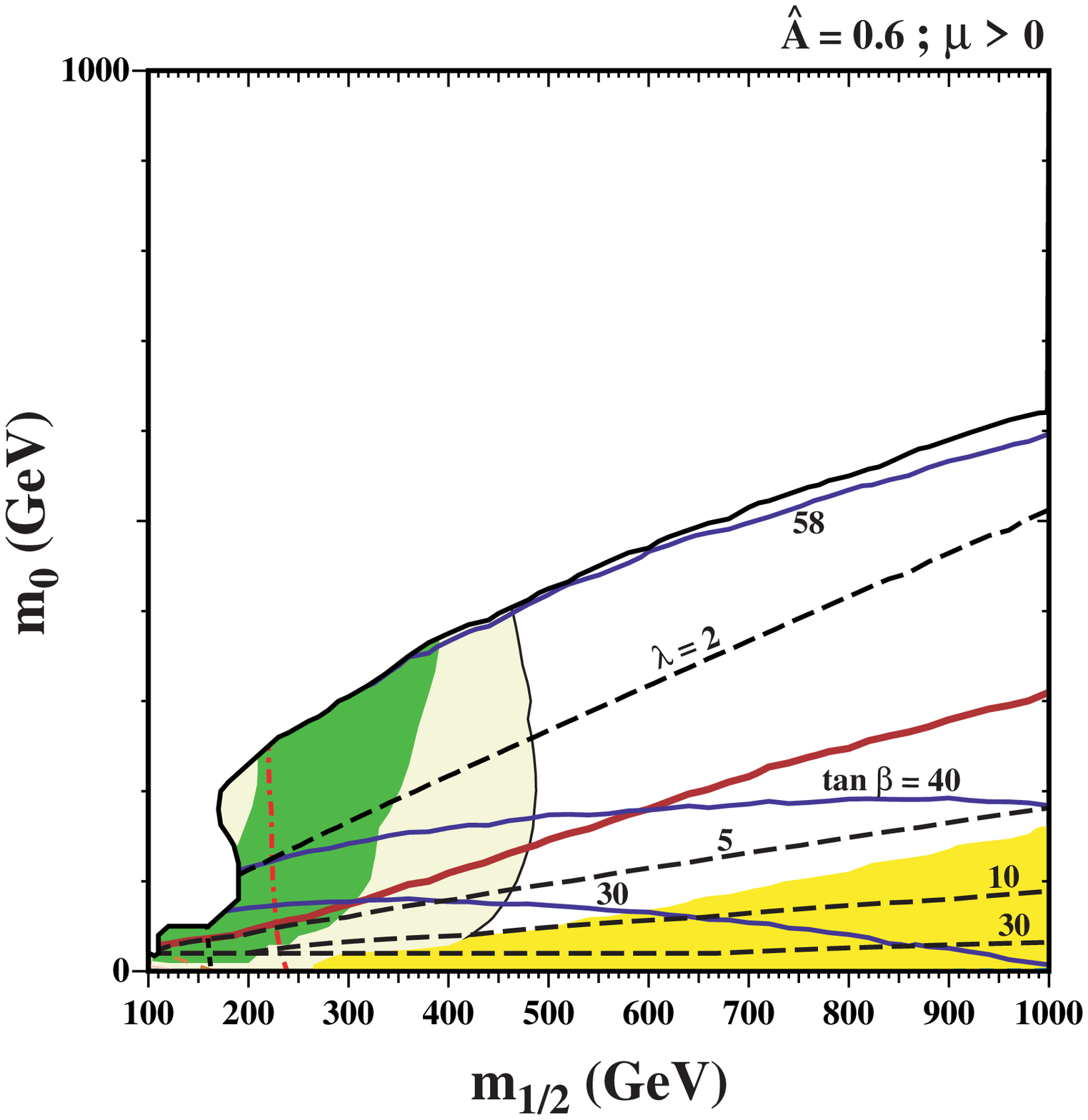,height=8cm}}
\mbox{\epsfig{file=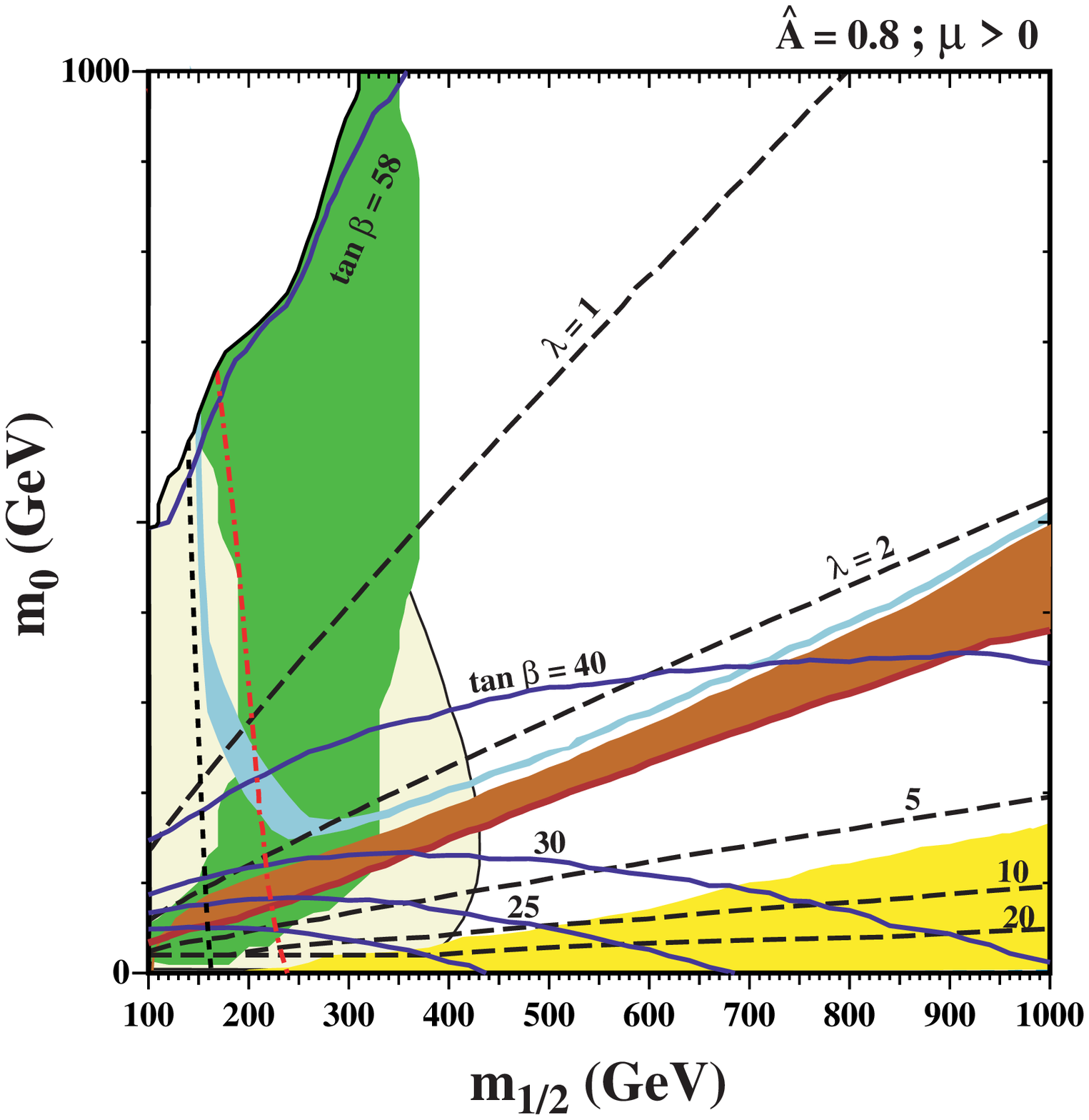,height=8cm}}
\end{center}
\begin{center}
\mbox{\epsfig{file=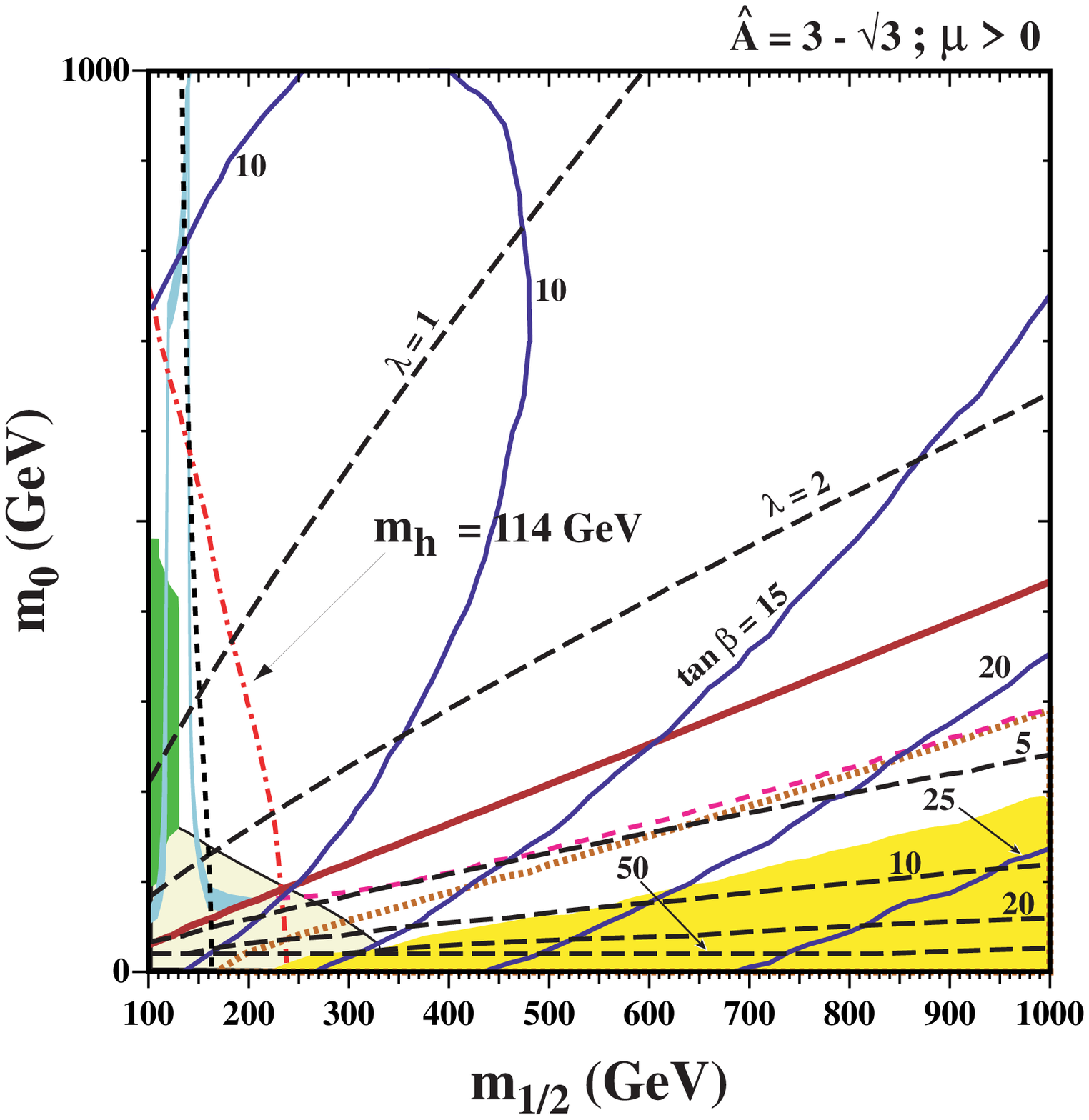,height=8cm}}
\mbox{\epsfig{file=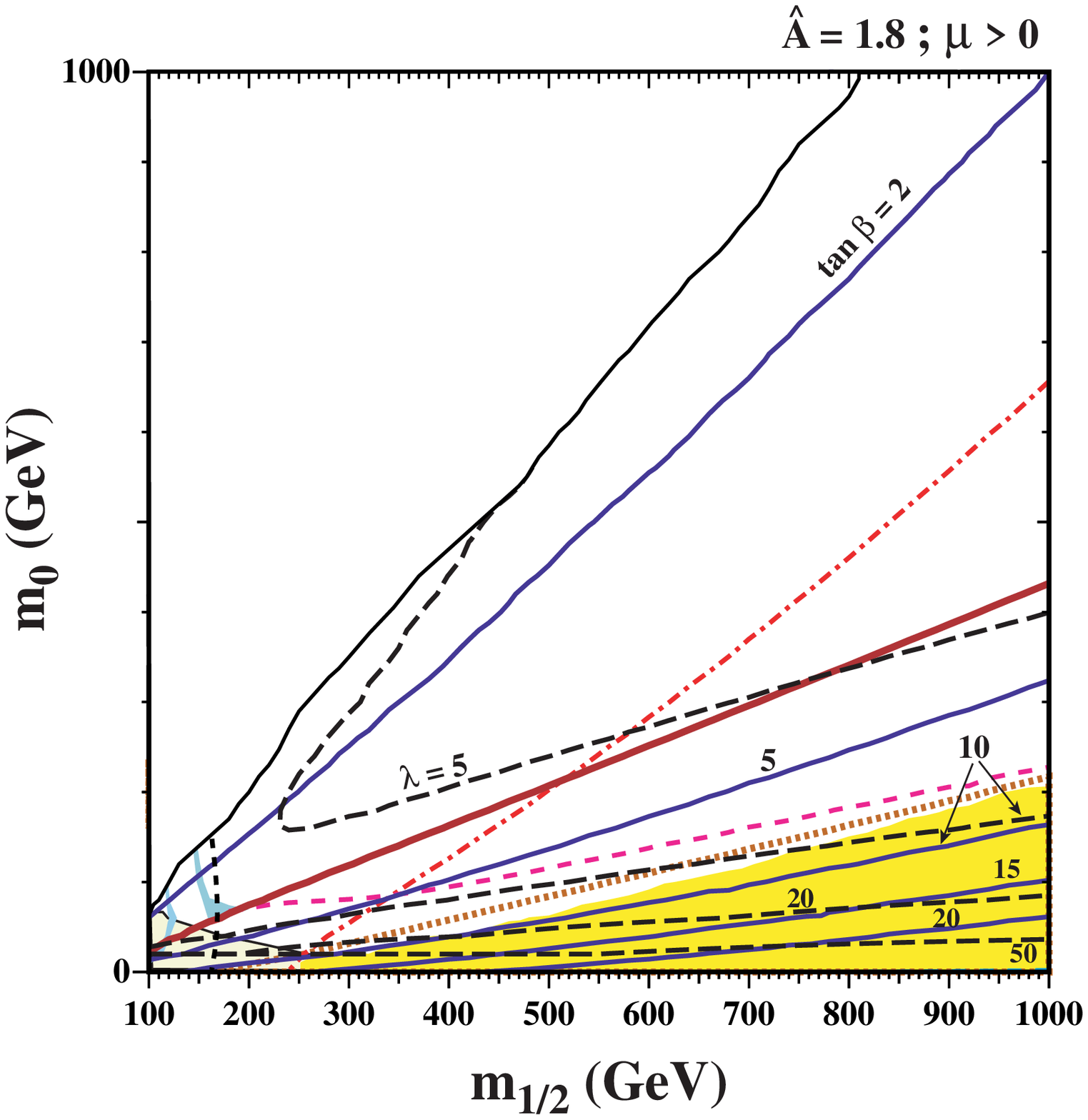,height=8cm}}
\end{center}
\caption{\it
Examples of $(m_{1/2}, m_0)$ planes in the Giudice-Masiero model, with 
contours of $\tan \beta$ 
superposed, for $\mu > 0$ and (a) ${\hat A} = 0.6$, (b) ${\hat A} = 0.8$, 
(c) ${\hat A} = 3-\sqrt{3}$ and (d) ${\hat A} = 1.8$.
In each panel, we show the regions excluded by 
the LEP lower limits on MSSM particles, those ruled out by $b
\to s \gamma$ decay~\protect\cite{bsg} (medium, green shading), and those 
disfavoured by $g_\mu - 2$ (very light, yellow shading). As before, a 
dotted 
(red) line shows where $m_\chi = m_{\tilde \tau}$ and the 
gravitino LSP region is bounded by a solid (chocolate) line in panels (a, 
c, d). The dark (red) wedge in panel (b) has a ${\tilde \tau}_1$ LSP 
and hence is disallowed. Within 
the gravitino LSP region, the relic density constraint is indicated by a 
dashed (pink) line and the BBN/CMB constraint on NSP decay by light 
(yellow) 
shading. The dashed black lines are contours of the Giudice-Masiero 
parameter $\lambda$.} 
\label{fig:GMp} 
\end{figure}

We display in Fig.~\ref{fig:GMp} some typical $(m_{1/2}, m_0)$ planes in
the Giudice-Masiero model for positive $\mu$. As in the previous minimal
supergravity VCMSSMs, we find no consistent electroweak solutions for
values of ${\hat A}$ much outside the range of values exhibited. In the
examples shown, there are no solutions above the topmost solid lines in
panels (a) for ${\hat A} = 0.6$ and (d) for ${\hat A} = 1.8$. For $\hat A <
-0.6$, $m_0 \lappeq 150$~GeV for $m_{1/2} \leq 1000$~GeV. Similarly for $\hat A
\gappeq 2.6$ only a small corner of the plane admits solutions. 

In panel (a) for ${\hat A} = 0.6$, corresponding to ${\hat B} = 0.75$,
there is no allowed region above the solid (chocolate) gravitino LSP line. 
Below this line,
we see an allowed region for $\tan \beta \gappeq 22$. However, we also
note that the corresponding values of $\lambda$ are quite large, $\lambda
\gappeq 5$. The situation is somewhat different for the case ${\hat A} =
0.8$, corresponding to ${\hat B} \simeq 0.64$, shown in panel (b) of
Fig.~\ref{fig:GMp}. In this case, we see that there is a narrow allowed
region along the $\chi - {\tilde \tau}$ coannihilation strip in the
neutralino LSP region for $\tan \beta \gappeq 33.5$, or $\gappeq 35$ if the
$g_\mu - 2$ constraint is taken into account. This region requires
$\lambda \gappeq 2$, which is relatively palatable. At lower $m_0$, there
is a disallowed dark (red) wedge where the ${\tilde \tau}_1$ is the 
LSP, and
below that a region where the gravitino is the LSP. The latter contains a
domain allowed by the BBN/CMB decay constraint, that appears for $\tan
\beta \gappeq 18$, or $\gappeq 20$ if one includes the 
$g_\mu - 2$ constraint. However, this region again has $\lambda \gappeq 5$.
Turning now to the Polonyi case ${\hat A} = 3 - \sqrt{3}$ shown in panel
(c) of Fig.~\ref{fig:GMp}, corresponding to ${\hat B} \simeq 0.27$, we see
that there is no allowed area in the neutralino LSP region above the 
dark solid (chocolate) line, and that the
allowed region in the gravitino LSP region requires $\tan \beta \gappeq
13.5$ and again $\lambda \gappeq 5$.  Similar features are seen in panel (d)
for ${\hat A} = 1.8$, corresponding to ${\hat B} = - 0.5$, where the only
allowed area - in the gravitino LSP region - requires even larger values
of $\lambda$ than the previous cases.

\begin{figure}
\begin{center}
\mbox{\epsfig{file=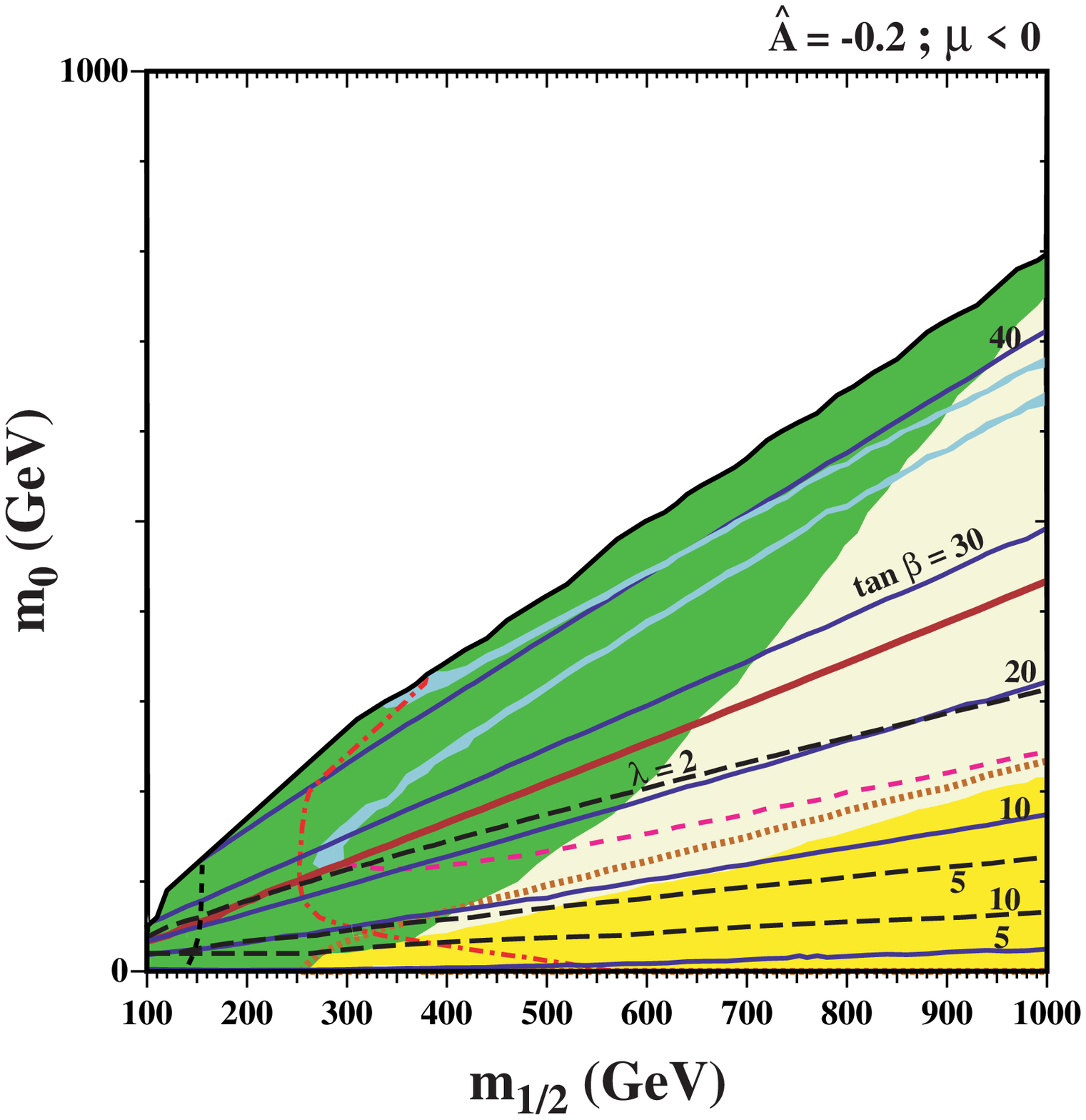,height=8cm}}
\mbox{\epsfig{file=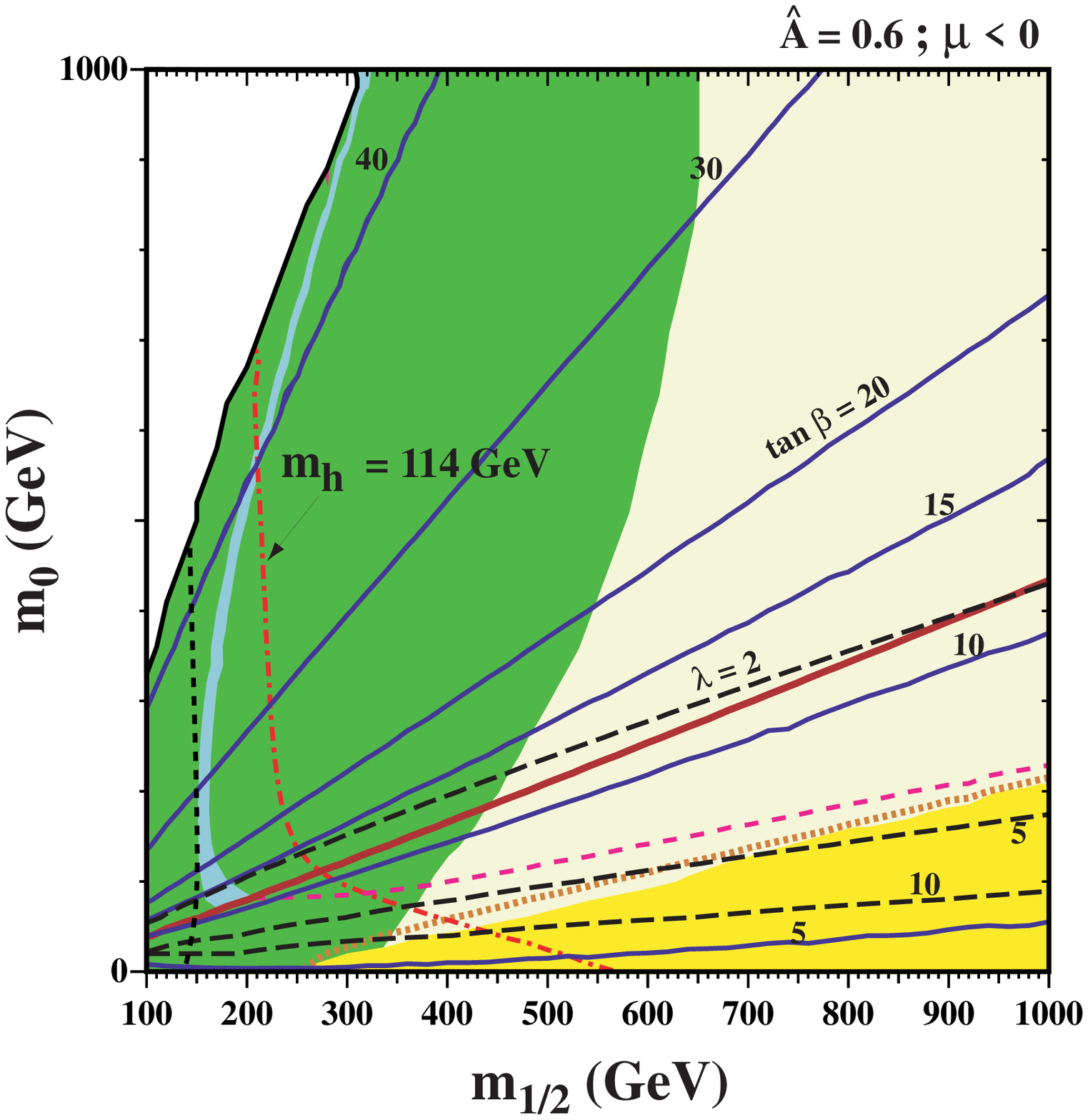,height=8cm}}
\end{center}
\begin{center}
\mbox{\epsfig{file=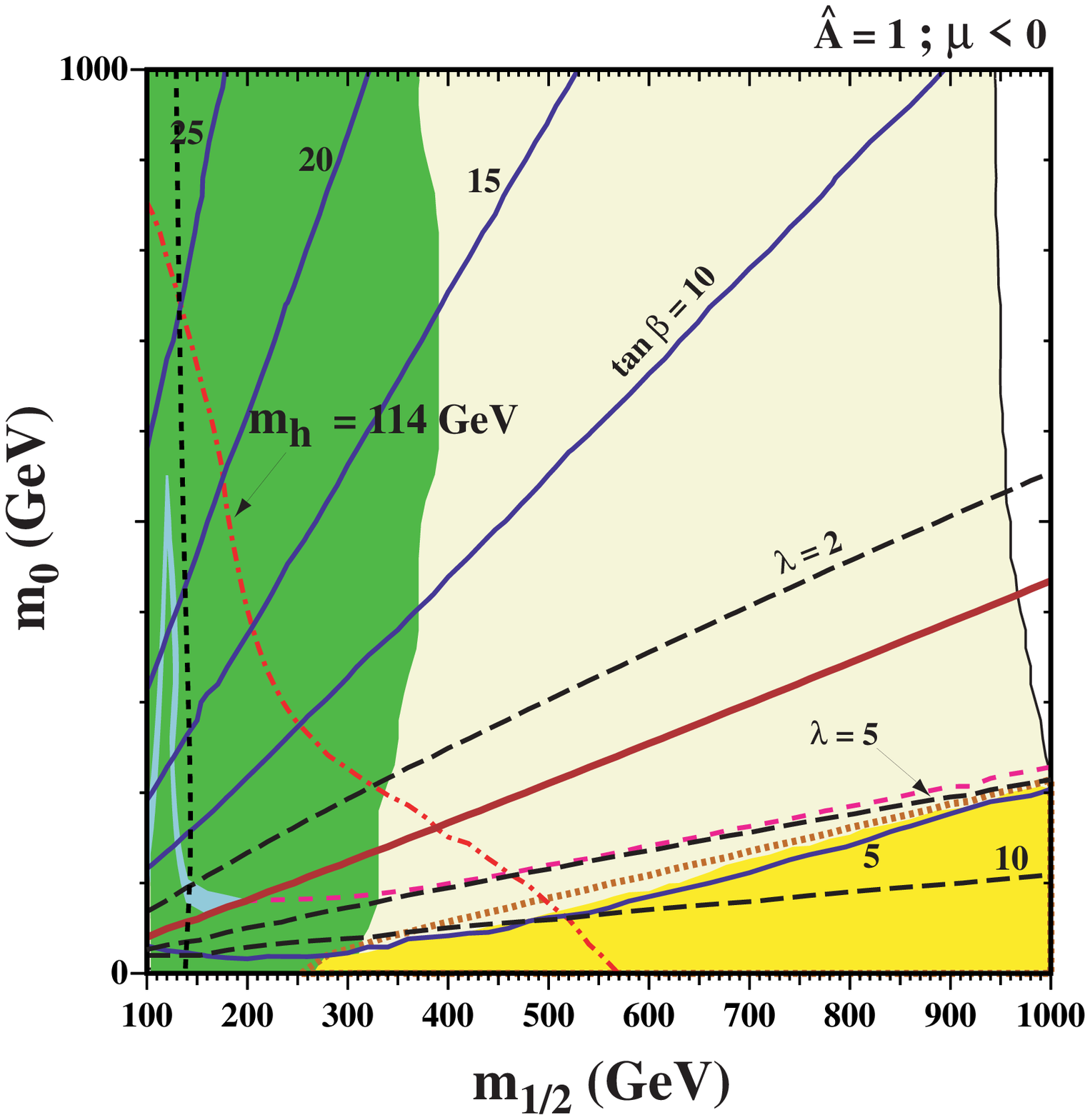,height=8cm}}
\mbox{\epsfig{file=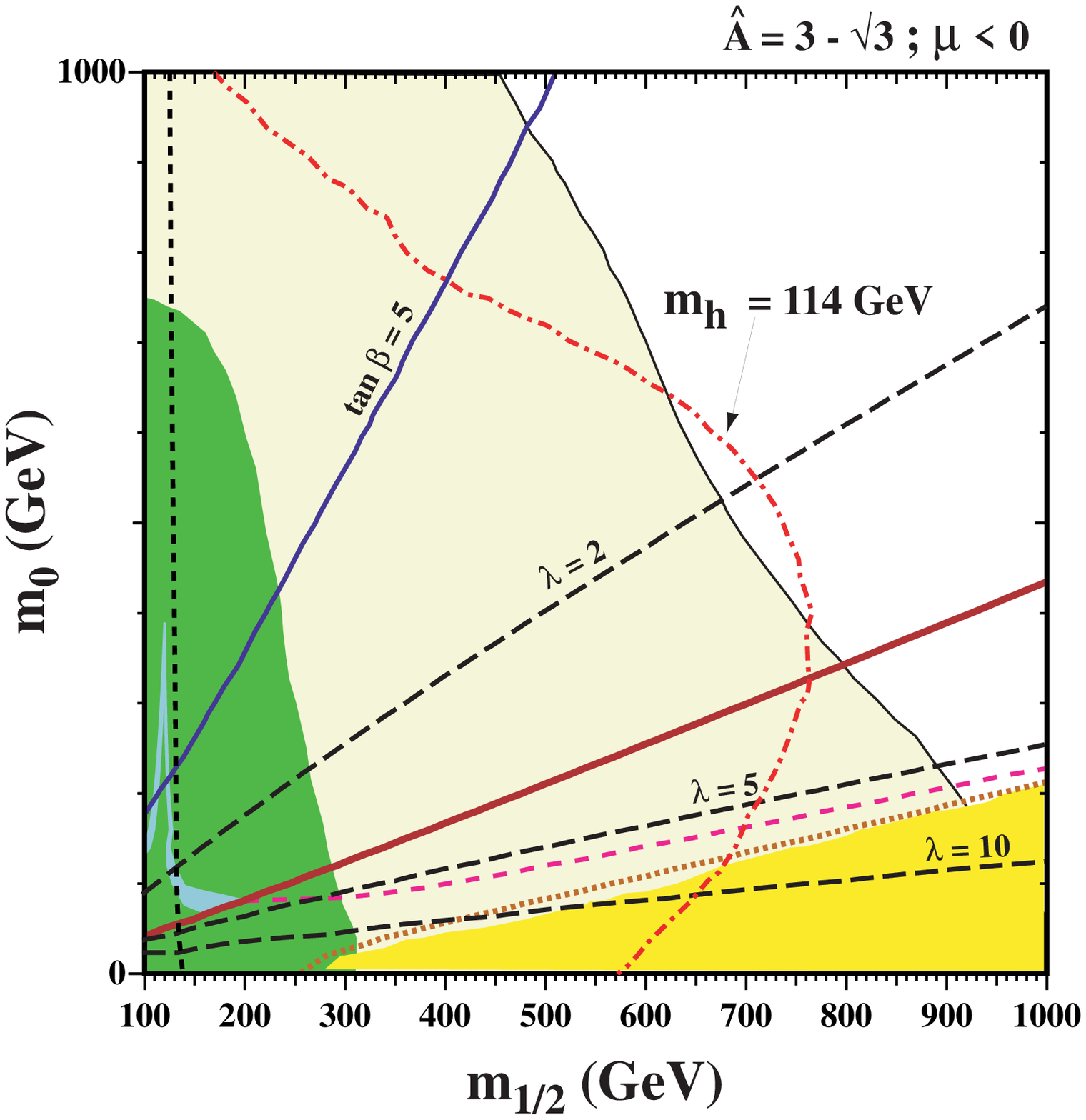,height=8cm}}
\end{center}
\caption{\it
Same as Fig.~\ref{fig:GMp}, but for $\mu < 0$ and (a) ${\hat A} = -0.2$, 
(b)
${\hat A} = 0.6$, (c) ${\hat A} = 1$, and (d) ${\hat A} = 3 - \sqrt{3}$.}
\label{fig:GMn}
\end{figure}

Fig.~\ref{fig:GMn} shows some analogous cases for $\mu < 0$. As before,
there are no consistent electroweak vacuum solutions for values of ${\hat
A}$ substantially outside the range of values shown, and none above the
topmost solid lines in panels (a) and (b). Panel (a) is for ${\hat A} =
-0.2$, corresponding to ${\hat B} \simeq 1.06$. It has two narrow strips
in the neutralino LSP region that are allowed if one discards the $g_\mu -
2$ constraint, appearing for $m_{1/2} \gappeq 800$~GeV and $m_0 \gappeq
500$~GeV for $\tan \beta > 38$ and $\lambda < 1.5$. Down in the gravitino
LSP area, there is a second allowed region with $\tan \beta \lappeq 13$
and $\lambda \gappeq 3$. For smaller values of $\hat A$, the allowed parameter
space is further squeezed. For example, for $\hat A = -1$ we find $m_0 
\lappeq
500$~GeV for $m_{1/2} \leq 1000$~GeV. In panel (b) for ${\hat A} = 0.6$,
corresponding
to ${\hat B} = 0.75$, the allowed neutralino LSP region has disappeared,
but a gravitino LSP region remains. Similar features are seen in panels
(c) and (d) for ${\hat A} = 1$ (${\hat B} = 0.5$) and ${\hat A} =
3-\sqrt{3}$ (${\hat B} \simeq 0.27$), respectively. For $\hat A \gappeq 2$,
solutions exist only in a small portion of the plane.

\section{Conclusions}

We have discussed in this paper the impacts of the theoretical,
experimental and cosmological constraints on some classes of VCMSSMs,
including minimal supergravity models and the Giudice-Masiero model. We
have presented unified treatments of the regions of parameter space in
these models where the LSP is a neutralino or the gravitino.

We have emphasized that the predictions of these models differ
significantly from those of the CMSSM. In particular, the CMSSM is
distinct from minimal supergravity: the former does not necessarily
require a fixed relation between the trilinear and bilinear soft
supersymmetry-breaking parameters $A, B$, nor equality between $m_0$ and
$m_{3/2}$, as required in minimal supergravity models. The values of $B$
required in generic realizations of the CMSSM generally bear no relation
to the values that would be derived in minimal supergravity models.

In addition to minimal supergravity models, we have discussed the simplest 
variant of the Giudice-Masiero model, which makes a brave attempt to 
provide a framework for calculating the Higgs-mixing superpotential 
term $\mu$.

There are a couple of striking features of these specific analyses that we 
note. One is that the range of $A$ is often very restricted: beyond this 
range, it is impossible to find consistent solutions to the electroweak 
vacuum conditions. 

A second observation is that, in both minimal supergravity and the
Giudice-Masiero model, a neutralino LSP is completely
excluded in many instances, and
the gravitino LSP regions are generally much more
extensive than the neutralino LSP regions. 
To some extent, this was to be
expected, since we impose the cosmological dark matter density and NSP
decay constraints on gravitino dark matter as one-sided upper limits,
rather than as narrow WMAP ranges as for the dark matter density 
constraint on
neutralino dark matter. This is because, in the case of gravitino dark
matter, the narrow range could be reached by postulating thermal gravitino
production with a suitable reheating temperature~\cite{buchmuller}. Of course,
in either the
neutralino or gravitino case, one could always postulate a supplementary
source of cold dark matter. In the case of neutralino dark matter, this 
possibility would broaden the WMAP density strip down to the $m_\chi = 
m_{\tilde \tau}$ boundary. However, the gravitino dark matter region would 
still, for many choices of the other supersymmetric parameters, occupy a 
larger area of the $(m_{1/2}, m_0)$ plane.

In any complete supersymmetric theory, one expects some relations between
supersymmetry breaking parameters, perhaps of the type discussed here. In
this case, some VCMSSM should be responsible for the low-energy sparticle
spectrum. However, we do not yet know what specific constraints are handed
down from the unification or string scales.  As we have emphasized in this
paper, the predictions in such models may differ greatly from those of the
more relaxed CMSSM and, {\it a priori}, those of a more general MSSM.
Analogous differences are also to be expected in the predicted cross
sections for direct and indirect searches for supersymmetric dark matter,
a topic we will consider elsewhere.

\vskip 0.5in
\vbox{
\noindent{ {\bf Acknowledgments} } \\
\noindent  The work of K.A.O., Y.S., and V.C.S. was supported in part
by DOE grant DE--FG02--94ER--40823. Y.S. would like to thank D.A. Demir 
for asking about the Giudice-Maisero mechanism in this context, and for 
helpful conversations.}

\end{document}